\documentclass[11pt]{article}
\usepackage{jcappub}
\usepackage{mathtools}
\usepackage{bm}
\usepackage{dsfont}
\usepackage{color}
\usepackage{array}
\usepackage{graphicx}
\usepackage{soul}
\usepackage{multirow}
\usepackage{multicol}
\usepackage{float}
\usepackage{hhline}
\usepackage[dvipsnames]{xcolor}
\usepackage[normalem]{ulem}
\usepackage{mathrsfs}
\usepackage{wasysym}
\usepackage[mathscr]{euscript}

\title{Proton Capture in Compact Dark Stars and Observable Implications}

\author[a,b]{Boris Betancourt Kamenetskaia,}
\author[a,b]{Anja Brenner,}
\author[a]{Alejandro Ibarra,}
\author[c]{Chris Kouvaris}
\affiliation[a]{Physik-Department, Technische Universit\"at M\"unchen, James-Franck-Stra\ss{}e, 85748 Garching, Germany}
\affiliation[b]{\normalsize Max-Planck-Institut f\"ur Physik (Werner-Heisenberg-Institut), F\"ohringer Ring 6, 80805 M\"unchen, Germany}
\affiliation[c]{Physics Division, National Technical University of Athens, 15780 Zografou Campus, Athens, Greece}

\emailAdd{boris.betancourt@tum.de}
\emailAdd{anja.brenner@tum.de}
\emailAdd{ibarra@tum.de}
\emailAdd{kouvaris@mail.ntua.gr}

\abstract{Asymmetric dark matter under certain conditions could form compact star-like objects, which can be searched either through gravitational lensing  or by observation of gravitational waves from binaries involving such compact objects. In this paper we analyze  possible signatures of such dark stars made of asymmetric dark matter with a portal to the Standard Model. We argue that compact dark stars could capture protons and electrons from the interstellar medium, which then accumulate in the core of the dark star, forming a very hot gas that emits X-rays or $\gamma$-rays. For dark matter parameters compatible with current laboratory constraints, compact dark stars could be sufficiently luminous to be detected at the Earth as point sources in the X-ray or $\gamma$-ray sky. 
}

\begin{document}
	\maketitle
	\flushbottom
	
	\section{Introduction} \label{sec:intro}

	Despite the overwhelming evidence for the existence of dark matter in galaxies, clusters of galaxies and the Universe at large scale, relatively little is known about the characteristics of the dark matter particle, such as its mass, spin, lifetime or its interactions with itself, with other particles of the dark sector, or with the particles of the observable sector (for reviews, see {\it e.g.} \cite{Jungman:1995df,Bertone:2004pz,Bergstrom:2000pn,Feng:2010gw}). Many dark matter searches,  motivated by the preeminence of the lightest neutralino as a dark matter candidate in the 1980s~\cite{Ellis:1983ew}, assume that the dark matter is an absolutely stable self-conjugated particle, with tiny self-interactions, and with interactions to the Standard Model  with strength comparable to the weak interaction (or slightly smaller, in order to avoid experimental constraints). On the other hand, from a phenomenological standpoint, these assumptions seem too restrictive. In fact, in recent years many scenarios have been constructed where these assumptions are relaxed. 
	
	In this paper we focus on scenarios where the dark matter particle is distinguishable from its antiparticle, due to a conserved ``dark charge" (analogous to the electric charge, or the baryon number). We will also assume that in the early Universe there was an excess of dark matter particles over antiparticles, so that most of the dark matter antiparticles annihilated at early times, yielding at current times a dark matter population of stable particles that cannot self-annihilate, due to the conservation of the dark matter charge. This scenario is commonly known as ``asymmetric dark matter", amounting to a dark sector analog to the proton-antiproton asymmetry in the visible sector \cite{Nussinov:1985xr,Barr:1990ca,Gudnason:2006yj} (for reviews, see \cite{Zurek:2013wia,Petraki:2013wwa}).
	
	In this class of scenarios, one may expect that a fraction of the dark matter could be in the form of ``dark stars", in the same way that a fraction of the protons in the galaxy form stars. This possibility was first discussed by Kaup at the end of the 1960s \cite{1968PhRv..172.1331K}. He considered a system of non-interacting complex scalars, bound by gravity, and supported against gravitational collapse by the Heisenberg uncertainty principle. The analysis was extended by Colpi et al. to the case where the scalar field has repulsive self-interactions \cite{Colpi:1986ye}. Asymmetric fermionic dark matter with attractive and repulsive self-interactions was first studied in \cite{Kouvaris:2015rea}. The case of asymmetric bosonic dark matter with attractive and repulsive self-interactions was studied in \cite{Eby:2015hsq}. Depending on the strength of the self-interaction and the mass of the dark matter particle, the aforementioned papers showed that there can be a very broad range of masses that such dark stars could have. Furthermore, it was demonstrated that such objects could have a large compactness and Love numbers that can distinguish mergers between them from others involving black holes or neutron stars~\cite{Maselli:2017vfi}.
	
	Depending on their mass, such compact dark stars can be searched in the sky through various methods including stellar microlensing~\cite{Macho:2000nvd,EROS-2:2006ryy,Niikura:2019kqi}, supernovae magnification~\cite{Zumalacarregui:2017qqd}, gravitational waves produced by dark star mergers~\cite{Maselli:2017vfi,Kavanagh:2018ggo,LIGOScientific:2019kan,Chen:2019irf}, dynamical constraints from wide binaries~\cite{Monroy-Rodriguez:2014ula} or dwarf galaxies~\cite{Brandt:2016aco}, or  dwarf galaxy heating~\cite{Lu:2020bmd} (for a review of constraints, see~\cite{Green:2020jor}). Furthermore, asymmetric dark stars made of dark matter kinetically mixed with photons can, under certain conditions, convert emitting dark photons to ordinary photons, providing a significant source of luminosity with a spectrum distinctly different from that of ordinary stars~\cite{Maselli:2019ubs}.  Most of the above observations suggest that not all dark matter in the galaxy is in the form of compact dark stars, but at most 1-10\%, depending on the mass (and mass spectrum).
	
	In this paper we investigate the possibility that compact dark stars could lead to additional signatures in experiments, aside from their gravitational interactions. We will focus mostly on scenarios where the dark matter particle interacts with the proton, which is widely studied in the context of direct dark matter detection experiments. In this case, protons in the interstellar medium could be captured by the  dark star, in a process analogous to the capture of dark matter in the Sun~\cite{Silk:1985ax,Gould:1987ir}. At the same time, electrons would also be captured, in order to keep the dark star electrically neutral. The protons and electrons would quickly thermalize within the interior of the dark star, and form a hot gas, which would emit radiation that could be detected in X-ray or $\gamma$-ray telescopes. Furthermore, the ``dark charge" that stabilizes the asymmetric dark matter and prevents the annihilation of two dark matter particles could be mildly broken by one or by two units, thus allowing respectively the slow decay or annihilation of dark matter particles into Standard Model particles, that could also be detected at the Earth. 
	
	This work is organized as follows: In Section \ref{sec:Boson_Stars} we review the basic properties of a bosonic dark star. In Section
	\ref{sec:CaptureRate} we determine the capture rate of protons in a dark star. In Section \ref{sec:Emission} we argue that the captured protons will thermalize quickly with the dark star and, as such, will emit radiation according to thermal mechanisms. In Section \ref{sec:Thermal_evolution} we analyze the impact of the emission of radiation in the dark star's thermal evolution, with emphasis on its luminosity. In Section \ref{sec:Signals} we discuss the prospects of observation of dark stars through their emission of electromagnetic radiation. In Section \ref{sec:Annihilation_Decay} we consider dark star signals due to a possible violation of dark matter number. Finally, in Section \ref{sec:conclusions} we present our conclusions.

	\section{Bosonic compact dark stars} \label{sec:Boson_Stars}
	
	In this section we review the derivation of the dark star properties, following closely the work by Colpi et al.~\cite{Colpi:1986ye}. Henceforth, we work with natural units, $c=\hbar=k_B=1$. We assume that the dark matter of our Universe is composed by self-interacting complex scalar particles, $\phi$, with mass $m$ and self-coupling strength $\lambda>0$. The Lagrangian density in curved spacetime reads
	\begin{align}
		{\cal L}=\frac{1}{2} g^{\mu\nu}\nabla_\mu\phi^* \nabla_\nu\phi-\frac{1}{2}m^2|\phi|^2-\frac{\lambda}{4} |\phi|^4,
	\end{align}
	where $\nabla_\mu$ denotes the covariant derivative and $g_{\mu\nu}$ is the metric tensor. The field $\phi$ satisfies the Klein-Gordon equation:
	\begin{align} 
		g^{\mu\nu}\nabla_\mu \nabla_\nu\phi-m^2 \phi-\lambda|\phi|^2\phi=0,
		\label{eq:KleinGordon}
	\end{align}
	while the space metric satisfies the Einstein equations:
	\begin{align}
		R_{\mu\nu}-\frac{1}{2}g_{\mu\nu}R=\frac{8\pi}{\mathrm{M}_{\mathrm{Pl}}^{2}}
		T_{\mu\nu},
		\label{eq:Einstein}
	\end{align}
	with $\mathrm{M}_{\mathrm{Pl}}=1.2\times 10^{19}$ GeV the Planck mass, and $T_{\mu\nu}$ the energy-momentum tensor of the scalar field:
	\begin{align}
		T^{\mu}_{~\nu}=\frac{1}{2}g^{\mu\sigma}\left(\nabla_{\sigma}\phi^{*}\nabla_{\nu}\phi+\nabla_{\sigma}\phi\nabla_{\nu}\phi^{*}\right)-\frac{1}{2}\delta^{\mu}_{\,\,\,\nu}\left(g^{\alpha\sigma}\nabla_{\alpha}\phi^{*}\nabla_{\sigma}\phi-m^2|\phi|^2-\frac{\lambda}{2}|\phi|^4\right).
		\label{eq:Tmunu}
	\end{align} 
	To solve the coupled Einstein-Klein-Gordon equations we consider static, spherically symmetric solutions. The line element, expressed in Schwarzschild coordinates, has the form, 
	\begin{equation}\label{eq:line_element}
		ds^{2}=B(r)dt^2-A(r)dr^2-r^2(d\theta^2+\sin^{2}\theta d\varphi^2),
	\end{equation}
	while the scalar field reads
	\begin{equation}
		\phi(r,t)=\Phi(r)\mathrm{e}^{-i\omega t},
	\end{equation}
	with $\Phi(r)$ a real function. Substituting in Eqs. (\ref{eq:KleinGordon}, \ref{eq:Einstein}, \ref{eq:Tmunu}), and after some manipulations, one finds that the real functions $A(r)$, $B(r)$ and $\Phi(r)$ satisfy  
	\begin{align}
		& \frac{A'}{A^2x}+\frac{1}{x^2}\left(1-\frac{1}{A}\right) = \left(\frac{\Omega^2}{B}+1\right)\sigma^2+\frac{\Lambda}{2}\sigma^4+\frac{(\sigma')^2}{A}, \label{ColpiSystemA} \\
		& \frac{B'}{ABx}-\frac{1}{x^2}\left(1-\frac{1}{A}\right) = \left(\frac{\Omega^2}{B}-1\right)\sigma^2-\frac{\Lambda}{2}\sigma^4+\frac{(\sigma')^2}{A}, \label{ColpiSystemB} \\
		& \sigma''+\left(\frac{2}{x}+\frac{B'}{2B}-\frac{A'}{2A}\right)\sigma'+A\left[\left(\frac{\Omega^2}{B}-1\right)\sigma-\Lambda\sigma^3\right] = 0, \label{eq:System}
	\end{align}
	where we have introduced the dimensionless quantities
	\begin{equation} \label{eq:Rescaling}
		\begin{aligned}
			&x=mr, && \sigma=\sqrt{4\pi} \frac{\Phi}{\mathrm{M}_{\mathrm{Pl}}}, &&  \Omega=\frac{\omega}{m}, && \Lambda=\frac{\lambda}{4\pi} \frac{\mathrm{M}_{\mathrm{Pl}}^2}{m^2},
		\end{aligned}
	\end{equation}
	and the prime denotes the derivative with respect to $x$. Note that Eq. (\ref{eq:System}) is obtained by making use of the conservation equation $\nabla_{\mu}T^{\mu\nu}=0$ for $\nu=1$ and a meaningful solution requires $\sigma'(0)=0$ so that  the second term in this equation remains finite.
	For the interpretation of the solutions, it is convenient to replace $A(x)$ by a ``mass function" $\mathcal{M}(x)$, defined as
	\begin{equation}
		A(x)=\left[1-\frac{2\mathcal{M}(x)}{x}\right]^{-1},
	\end{equation}
	which, using Eq.~(\ref{ColpiSystemA}), satisfies
	\begin{equation}\label{MassParameter}
		\mathcal{M}'(x)=x^2\left[\frac{1}{2}\left(\frac{\Omega^2}{B}+1\right)\sigma^2+\frac{\Lambda}{4}\sigma^4+\frac{1}{2}\frac{(\sigma')^2}{A}\right].
	\end{equation}
	Furthermore, in order to simplify the form of the equations, it is convenient to introduce the rescaled quantities
	\begin{equation}
		\begin{aligned}
			&x_{*}=x\Lambda^{-\frac{1}{2}}, && \sigma_{*}=\sigma\Lambda^{\frac{1}{2}}, && \mathcal{M}_{*} =\mathcal{M}\Lambda^{-\frac{1}{2}},
		\end{aligned}
	\end{equation}
	so that Eqs.~(\ref{ColpiSystemB}, \ref{eq:System}, \ref{MassParameter}) take the form
	\begin{align}
		&\mathcal{M}_{*}'(x_{*})=x_{*}^2\left[\frac{1}{2}\left(\frac{\Omega^2}{B}+1\right)\sigma_{*}^2+\frac{1}{4}\sigma_{*}^4+\frac{1}{2\Lambda}\frac{(\sigma')^2}{A}\right], \label{ColpiSystemRedefA} \\
		&\frac{B'}{ABx_{*}}-\frac{1}{x_{*}^2}\left(1-\frac{1}{A}\right) = \left(\frac{\Omega^2}{B}-1\right)\sigma_{*}^2-\frac{1}{2}\sigma_{*}^4+\frac{1}{\Lambda}\frac{(\sigma_{*}')^2}{A}, \label{ColpiSystemRedefB}\\
		&\frac{1}{\Lambda}\sigma_{*}''+\frac{1}{\sqrt{\Lambda}}\left(\frac{1}{\sqrt{\Lambda}}\frac{2}{x_{*}}+\frac{B'}{2B}-\frac{A'}{2A}\right)\sigma_{*}'+A\left[\left(\frac{\Omega^2}{B}-1\right)\sigma_{*}-\sigma_{*}^3\right]=0,
	\end{align}
	where now the prime denotes the derivative with respect to $x_*$. Here $A$ is a function of ${\cal M}_*$, therefore the independent variables in this set of differential equations are ${\cal M}_*$, $B_*$ and $\sigma_*$.
	From a Particle Physics standpoint, it is plausible that $\lambda\leq {\cal O}(1)$ and $m\ll \mathrm{M}_{\mathrm{Pl}}$, which implies $\Lambda\gg 1$, {\it cf.} Eq.~(\ref{eq:Rescaling}). In this limit, the above equations simplify to
	\begin{subequations}\label{ColpiSystemHighLam}
		\begin{align}
			& \mathcal{M}_{*}'=4\pi x_{*}^2\rho_{*}\,, \label{ColpiSystemHighLamA} \\
			& \frac{B_{*}'}{AB_{*}x_{*}}-\frac{1}{x_{*}^2}\left(1-\frac{1}{A}\right) = 8\pi p_{*}\,, \label{ColpiSystemHighLamB}\\
			& \sigma_{*}=\left(\frac{1}{B_*}-1\right)^{\frac{1}{2}}, 
		\end{align}
	\end{subequations}
	where we have introduced 
	\begin{subequations}
		\begin{align}
			&\rho_{*}=\frac{1}{16\pi}\left(\frac{3}{B_{*}}+1\right)\left(\frac{1}{B_{*}}-1\right), \label{rhoStar} \\
			&p_{*}=\frac{1}{16\pi}\left(\frac{1}{B_{*}}-1\right)^2, \label{pStar} \\
			&B_{*}=\frac{B}{\Omega^2}. \label{BStar}
		\end{align}
	\end{subequations}
	Here, $\rho_*$ can be clearly interpreted as a dimensionless mass density distribution, expressed in terms of the  variable $x_*$. To interpret $p_*$, we calculate the derivative of $p_*$ with respect to the rescaled radial coordinate $x_*$. After some algebraic manipulations, one obtains:
	\begin{equation}\label{OV Equation Dimensionless}
		\frac{dp_{*}}{dx_{*}}=-\frac{1}{x_{*}^2}\left(4\pi p_{*}x_{*}^3+\mathcal{M}_{*}\right)\left(\rho_{*}+p_{*}\right)\left(1-\frac{2\mathcal{M}_{*}}{x_{*}}\right)^{-1},
	\end{equation}
	which has the form of the Tolman-Oppenheimer-Volkoff equation. Therefore, $p_*$ is the dimensionless pressure inside the dark star. Furthermore, from Eqs. (\ref{rhoStar}, \ref{pStar}) one obtains the equation of state 
	\begin{equation}\label{DimensionlessEoS}
		p_{*}=\frac{1}{36\pi}\left(\sqrt{1+12\pi\rho_{*}}-1\right)^2.
	\end{equation}
	The system of equations Eq.~(\ref{ColpiSystemHighLam}) expressed in terms of  ${\cal M}_*$, $B_*$ and $\sigma_*$, can be recast in terms of the more familiar quantities ${\cal M}_*$, $\rho_*$ and $p_*$ as
	\begin{subequations}\label{eq:DSStellar Structure}
		\begin{align}
			& \frac{d\mathcal{M}_{*}}{dx_{*}}=4\pi x_{*}^2\rho_{*}, \label{DSStellarStructureA} \\
			&\frac{dp_{*}}{dx_{*}}=-\frac{1}{x_{*}^2}\left(4\pi p_{*}x_{*}^3+\mathcal{M}_{*}\right)\left(\rho_{*}+p_{*}\right)\left(1-\frac{2\mathcal{M}_{*}}{x_{*}}\right)^{-1}, \\
			& p_{*}(\rho_{*})=\frac{1}{36\pi}\left[\sqrt{1+12\pi\rho_{*}}-1\right]^2. \label{DSStellarStructureC}
		\end{align}
	\end{subequations}
	We solve these equations using as boundary conditions that the total mass at the core is ${\cal M}_*(0)=0$ and  that the (normalized) pressure at the core is $p_*(0)=p_{*,c}$. Alternatively, using Eq.~(\ref{DimensionlessEoS}), one can use as boundary condition the (normalized) density at the core,  $\rho_*(0)=\rho_{*,c}$,  which ultimately depends on the environment where the dark star was formed. Finally, we impose that the pressure (or density) at the dark star's surface vanishes, that is,  $p_*(R_*)=0$ ($\rho_*(R_*)=0$), which allows us to implicitly define the dark star's radius $R_*$.
	
\begin{figure}[t!]
		\centering
		\includegraphics[width=.48\textwidth]{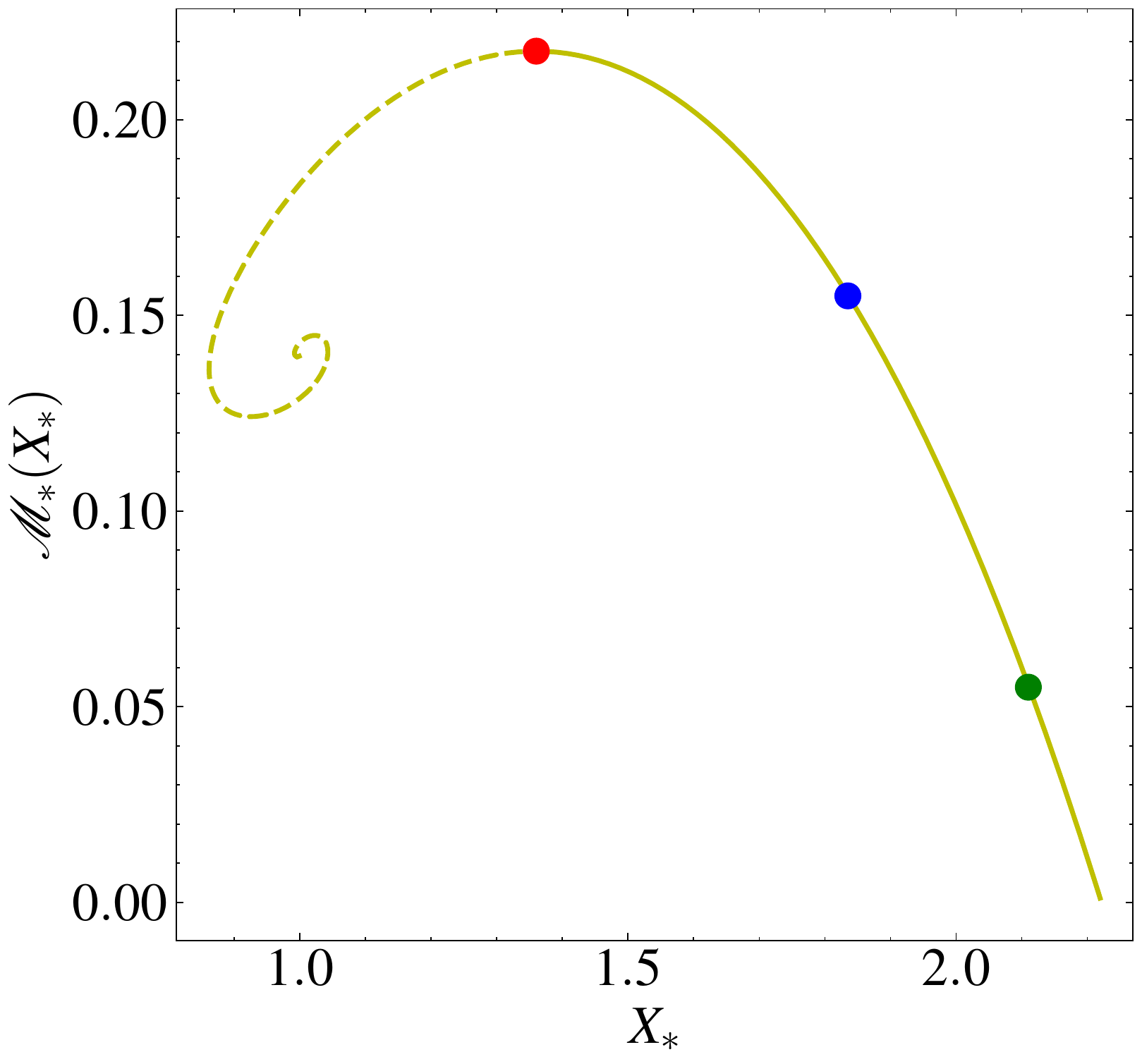}
		\includegraphics[width=.48\textwidth]{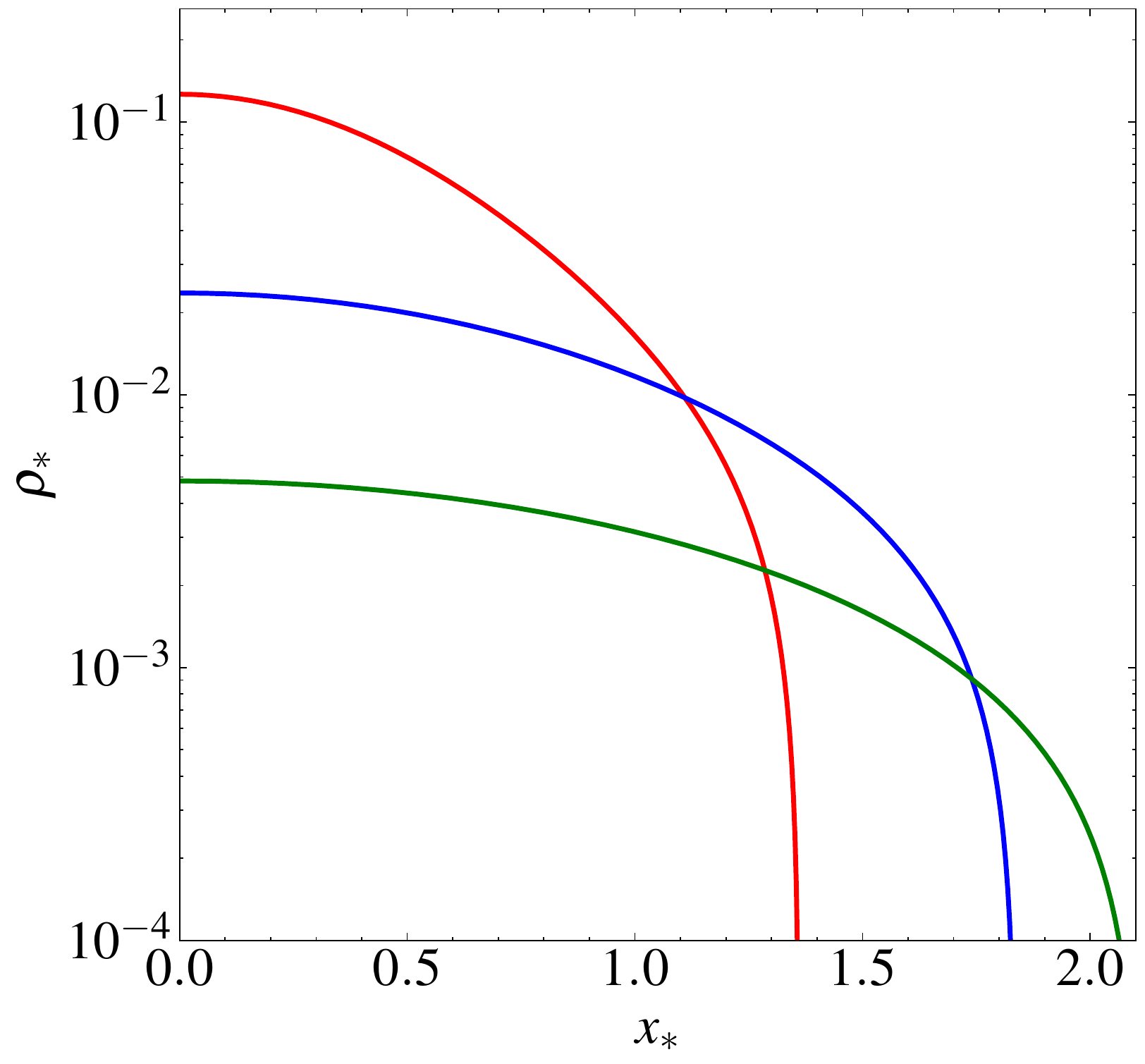}
		\caption{Left panel: Mass versus radius of the bosonic dark star. Right panel: Normalized density profile for the points indicated in the left panel.}
		\label{fig:mass_vs_x}
	\end{figure}

	We show in the left panel of Fig. \ref{fig:mass_vs_x}, the relation between the normalized mass and radius of the bosonic dark star; the solid and dashed line indicate whether the hydrostatic equilibrium inside the dark star is stable or unstable respectively. Different points on the curve correspond to different core densities. The red point corresponds to $\rho_{*,c}=0.126$, and leads to the largest possible mass of the dark star. The density profile associated to this choice is shown in the right panel of the Figure in red. As the mass decreases, the core density decreases too, while the size of the dark star increases, as illustrated in the Figure for the specific choices $\rho_{*,c}=0.0236$ (blue), $\rho_{*,c}=0.048$ (green).

	In order to understand the physical implications of these results it is convenient to undo the various changes of variables in the derivation of Eq.~(\ref{eq:DSStellar Structure}), and  express the dimensionful mass, radius, density and pressure ($M$, $r$, $\rho$ and $p$) in terms of their dimensionless counterparts ($M_*$, $r_*$, $\rho_*$ and $p_*$). The result is:
	\begin{subequations} \label{eq:Unit Conversion}
		\begin{align} 
			& M=(0.46 M_\odot)\,\lambda^{\frac{1}{2}}\left(\frac{1\ \mathrm{GeV}}{m}\right)^2M_{*}, \label{Mass Conversion2} \\
			& r=(0.68\ \mathrm{km})\lambda^{\frac{1}{2}}\left(\frac{1\ \mathrm{GeV}}{m}\right)^2x_{*}, \label{Length Conversion2} \\
			&\rho=\left(2.9\times10^{21}\ \frac{\mathrm{kg}}{\mathrm{m}^3}\right)\lambda^{-1}\left(\frac{1\ \mathrm{GeV}}{m}\right)^{-4}\rho_{*} ,\label{Density Conversion} \\
			&p=\left(2.6\times10^{38}\ \mathrm{Pa}\right)\lambda^{-1}\left(\frac{1\ \mathrm{GeV}}{m}\right)^{-4}p_{*}, \label{Pressure Conversion} 
		\end{align}
	\end{subequations}
	where $M_\odot=2\times 10^{30}$ kg is the mass of the Sun. For $\lambda=1$ and $m=1$ GeV, one finds that the dark star has a mass of the order of $0.1 M_\odot$ and  radius of the order of 1 km, namely, they are very compact objects. We should emphasize here that self-interactions have a dominant effect on the density profile of these stars. In the absence of self-interactions, it is the uncertainty principle that can oppose to the collapse. In such a case, the maximum mass is expected to be $\sim M_{\text{Pl}}^2/m$, which means that in the previous example of a $m=1$ GeV mass, the maximum mass is roughly $10^{-18} M_\odot$, quite different from $0.1 M_\odot$.
	
	\section{Proton capture by a dark star}\label{sec:CaptureRate}
	
	We consider a population of protons with number density $n_p$ and isotropic velocity distribution $f(u)$ far away from the dark star. To calculate the capture rate of protons in the dark star, we closely follow  the formalism developed in \citep{Press:1985ug,Gould:1987ir} to determine the capture rate of dark matter particles in the Sun, adapted to incorporate relativistic effects (see also 
	\cite{Goldman:1989nd,Kouvaris:2007ay,Garani:2018kkd,Bell:2020jou}).
	
	Let us consider a sphere centered at the dark star of radius $R\gg R_{DS}$. The flux of particles going inwards across this surface reads
	\begin{equation} \label{eq:flux}
		dF=\frac{1}{4}n_p\,f(u)u\,du\,d\cos^2\alpha,
	\end{equation}
	where $\alpha$ is the angle between the proton position vector and its velocity at the surface of the sphere~\cite{2017PhRvD..96f3002B}. It is convenient to recast this expression in terms of constants of motion, which characterize the worldline of the proton. The geodesic equations of a massive particle in the geometry defined by Eq.~(\ref{eq:line_element}) can be written as
	\begin{subequations}
		\begin{align}
			&r^2\frac{d\phi}{dt}=JB(r), \label{phi_eq4} \\
			&\frac{A(r)}{B(r)^2}\left(\frac{dr}{dt}\right)^2+\frac{J^2}{r^2}-\frac{1}{B(r)}=-k, \label{r_eq4}  
		\end{align}
	\end{subequations} 
	where the constants $(1-k)/2$ and $J$ can be physically interpreted as the energy per unit mass (in the weak field limit) and angular momentum per unit mass, respectively (see e.g.~\cite{1972gcpa.book.....W} for details). Alternatively, the trajectory of the proton can be described by the orbit equation
	\begin{align}
		\frac{A}{B^2r^4}\left(\frac{dr}{d\phi}\right)^2+\frac{1}{r^2B^2}-\frac{1}{B^3J^2}=-\frac{k}{J^2B^2} \label{eq:dr_dphi}.
	\end{align}

	For a proton with impact parameter $b=R \sin\alpha$ and speed $u$, one finds $J= R u \sin\alpha$,  which allows to recast Eq.~(\ref{eq:flux}) in terms of $J$.  Setting $dr/d\phi=0$ in Eq.~(\ref{eq:dr_dphi}) and solving for $J$ gives the maximum angular momentum allowed for a particle to cross a shell of radius $r$,  $J_{\rm max}= r \sqrt{1/B(r)-k}$. Furthermore, to determine the constant $k$, we consider a proton  approaching the dark star from infinity with speed $u$. At $r\rightarrow \infty$, $dr/dt= -u$, $A,B= 1$, $J=ub$, therefore $k=1-u^2$.
	
	The total number of particles  entering the sphere per unit coordinate time of a stationary observer located at the distance $R$ is then calculated integrating the incoming flux Eq.~(\ref{eq:flux})  over the whole sphere, giving
	\begin{equation}\label{eq:incoming_rate}
		d\mathcal{F}=\frac{dN}{dt}=4\pi R^2dF=\pi n_p\frac{f(u)du}{u}dJ^2.
	\end{equation}
	
	To estimate the capture rate of protons by the dark star, we assume for simplicity that any particle that suffers a scattering will get gravitationally bound to the dark star and would eventually sink to the center. We expect this to be a good approximation, since the proton velocity at the location of the target is much larger than at infinity, due to the gravitational attraction of the dark star. Therefore, even if the proton loses a small fraction of its velocity in the scattering, its final velocity will be smaller than the escape velocity, and will become bound to the dark star. We would like to find the probability of a proton to scatter and be captured while traversing a shell of the star with width $dr$. Since in general the particle will cross the shell in an angle, the distance $dl$ through a spherical shell of the dark star at coordinate distance $r$ from the center is given by
	\begin{equation}
		P=\frac{dl}{l^{\rm shell}_{\rm mfp}},
	\end{equation}
	where $l^{\rm shell}_{\rm mfp}=\frac{1}{n^{\rm shell}(r)\sigma}$ is the mean free path at that shell. To calculate $dl$, we consider the line element at $dt=0$, $dl^2=-ds^2=A(r)dr^2+r^2d\phi^2$ or
	
	\begin{equation}
		dl=2\sqrt{A(r)}dr\sqrt{1+\frac{r^2}{A(r)\left(dr/d\phi\right)^2}},
	\end{equation}
	where the factor of $2$ takes into account the fact that the proton crosses a particular shell twice (one as it enters and one as it exits the star). By Eq. (\ref{eq:dr_dphi}) and the expression for $J_{\rm max}$, this length can be rewritten as
	
	\begin{equation}\label{eq:shell_distance}
		dl=\frac{2\sqrt{A(r)}dr}{\sqrt{1-\left(\frac{J}{J_{\rm max}}\right)^2}}.
	\end{equation}
	On the other hand, the mean free  path measured by an observer on the shell of the star is
	
	\begin{equation}\label{eq:shell_mfp}
		l^{\rm shell}_{\rm mfp}=\frac{dV^{\rm shell}}{dN \sigma}=\frac{\sqrt{A(r)}r^2\sin\theta drd\theta d\phi}{dN \sigma}=\frac{\sqrt{A(r)}dV^{\rm far}}{dN \sigma}=\sqrt{A(r)}l^{\rm far}_{\rm mfp}=\frac{\sqrt{A(r)}}{n(r)\sigma}.
	\end{equation}
	
	By use of Eqs. (\ref{eq:shell_distance}) and (\ref{eq:shell_mfp}), the capture probability then reads
	\begin{equation}\label{eq:capture_probability}
		P=\frac{2n(r)\sigma dr}{\sqrt{1-\left(\frac{B}{1-B+u^2B}\right)\frac{J^2}{r^2}}}.
	\end{equation}
	We can then estimate the differential capture rate at that shell from the incoming flux of protons with velocity at infinity $u$ and angular momentum $J$, Eq.~(\ref{eq:incoming_rate}), and the capture probability, Eq.~(\ref{eq:capture_probability}): 
	\begin{equation}
		dC=d\mathcal{F}\times P= \displaystyle{2\pi n_p\sigma n(r)dr\frac{f(u)du}{u}\frac{dJ^2}{\sqrt{1-\left(J/J_{\rm max}\right)^2}}}.
	\end{equation}
	Finally, performing the integration over angular momenta, velocities and radial coordinates, we obtain the total capture rate	
	\begin{equation}
		\label{cap1}
		C=n_p \sigma \int_{0}^{\infty}{\frac{f(u)du}{u}}\int_{0}^{R_{\rm DS}}{4\pi r^2 n(r)\left[\frac{1-B(r)}{B(r)}\right]dr},
	\end{equation}
	where we have disregarded the $u$ dependent terms of $J_{\rm max}$ which are negligible since the proton velocity away from the star is insignificant compared to its speed as it crosses the star. 
	Let us consider for concreteness a velocity distribution with a  Maxwell-Boltzmann form:
	\begin{equation}
		\label{dist}
		f(u)du=6\sqrt{\frac{3}{2\pi}}\frac{1}{\bar{v}^3}u^2\mathrm{e}^{-\frac{3}{2}\frac{u^2}{\bar{v}^2}}du,
	\end{equation}
	where $\bar{v}$ is the velocity dispersion . In this case,
	\begin{equation}
		\label{cap2}
		C=\frac{4\sqrt{6\pi}n_p \sigma}{\bar{v}}\int_{0}^{R_{\rm DS}}{\frac{1-B(r)}{B(r)}r^2n(r)dr},
	\end{equation}
	which agrees with \cite{PhysRevD.40.3221,Kouvaris:2007ay}. However, it differs slightly from the capture rate derived in~\cite{Bell:2020jou}. The difference is that within the integral over $r$ in Eq.~(\ref{cap2}), \cite{Bell:2020jou} has a factor $\sqrt{1-B(r)}/B(r)$ while in our calculation we find $(1-B(r))/B(r)$, which on basis of consistency with the previously stated articles, we believe is the correct term. We would like to note that both conflicting terms, $\sqrt{1-B(r)}$ and $1-B(r)$, only take values in the interval $[0,1]$, and the maximum difference between them (at $B=0.75$) is of $50\%$. This means that the numerical deviation between both solutions is at most only a multiplicative factor of ${\cal O}(1)$.
	
	This capture rate is constrained by the maximal value called the ``geometric limit'', which assumes that all incoming particles are captured and sets (\ref{eq:capture_probability}) equal to unity. In this case, the capture rate is just the integral over the incoming rate given by Eq. (\ref{eq:incoming_rate})
	
	\begin{equation}
		C_{\rm geo}=\int d\mathcal{F}(R_{\rm DS})=\pi n_p \int_{0}^{\infty}{\frac{f(u)du}{u}\int_{0}^{J_{\rm max}(R_{\rm DS})}{dJ^2}},
	\end{equation}
	with $J_{\rm max}(R_{\rm DS})\approx R_{\rm DS}\sqrt{\frac{1-B(R_{\rm DS})}{B(R_{\rm DS})}}$. The geometric cross section then reads
	
	\begin{equation}\label{C_geo1}
		C_{\rm geo}=\frac{\sqrt{6\pi}n_p R_{\rm DS}^2}{\bar{v}}\left(\frac{1-B(R_{\rm DS})}{B(R_{\rm DS})}\right).
	\end{equation}	
	Finally it is easy to generalize the capture rate in the case where the  star is moving. Boosting the star to nonzero velocity results in a new distribution (see e.g.~\cite{Gould:1987ir}) 
	
	\begin{equation}
		f_{v_s}(u)=f(u)e^{-\frac{3v_s^2}{2\bar{v}^2}}\frac{\sinh (3u v_s/\bar{v}^2)}{3u v_s/\bar{v}^2},   
	\end{equation}
	where $f(u)$ is given by Eq.~(\ref{dist}) and $v_s$ is the star velocity. Solving the velocity integral in Eq.~(\ref{cap1}) using the boosted distribution we get
	
	\begin{equation}
		C_{v_s}=\frac{4\pi n_p\sigma}{v_s} {\rm Erf}
		\left ( \sqrt{\frac{3}{2}}\frac{v_s}{\bar{v}} \right )\int_{0}^{R_{\rm DS}}{\frac{1-B(r)}{B(r)}r^2n(r)dr}.
	\end{equation}
	Note that by taking the limit $v_s\rightarrow 0$ is well behaved,  recovering Eq.~(\ref{cap2}). Evaluating for the red profile in Fig. \ref{fig:mass_vs_x}, we obtain the numerical estimate
	
	\begin{align}\label{eq:capture_num}
		C_{v_s}&=(1.3\times10^{15}\ \mathrm{s}^{-1})\left(\frac{m}{1\ \mathrm{GeV}}\right)^{-1}\left(\frac{R_{\rm DS}}{1\ \mathrm{km}}\right)^{3}\left(\frac{\rho_{\rm core}}{10^{18}\ \mathrm{gr}/\mathrm{cm}^3}\right)\cr
		&\times\left(\frac{\sigma}{10^{-50}\ \mathrm{cm}^2}\right)\left(\frac{n_p}{10^{-5}\ \mathrm{cm}^{-3}}\right)\left(\frac{v_s}{10^{-3}}\right)^{-1}\mathrm{Erf}\left[\sqrt{\frac32}\left(\frac{v_s}{\bar{v}}\right)\right].
	\end{align} 
	For the sake of completeness, the geometric capture rate with this generalization reads
	
	\begin{equation}
		C_{v_s,\rm geo}=\frac{\pi n_p R_{\rm DS}^2}{v_s} {\rm Erf}
		\left ( \sqrt{\frac{3}{2}}\frac{v_s}{\bar{v}} \right )\left(\frac{1-B(R_{\rm DS})}{B(R_{\rm DS})}\right),
	\end{equation}
	which reproduces Eq. (\ref{C_geo1}) as $v_s\rightarrow 0$, as expected. Numerically, in the same way as for Eq. (\ref{eq:capture_num}), the geometric rate is
	
	\begin{equation}\label{eq:capture_geo_num}
		C_{v_s,\rm geo}=(4.4\times10^{18}\ \mathrm{s}^{-1})\left(\frac{R_{\rm DS}}{1\ \mathrm{km}}\right)^{2}\left(\frac{n_p}{10^{-5}\ \mathrm{cm}^{-3}}\right)\left(\frac{v_s}{10^{-3}}\right)^{-1}\mathrm{Erf}\left[\sqrt{\frac32}\left(\frac{v_s}{\bar{v}}\right)\right].
	\end{equation}

	\section{Photon emission from captured protons}\label{sec:Emission}

	Once the protons get captured by the dark star, they interact repeatedly with the dark matter particles in the interior of said star, and quickly thermalize with it provided we are in the  appropriate boson mass and dark matter-proton cross section range. The thermalization process takes place in two stages: in the first one, protons that interact within the dark star lose sufficient energy to become gravitationally bound to it. Subsequently, they might get out of the star but are doomed to return passing through the star over and over. In each passing there is substantial probability of rescattering and losing more energy, eventually having orbits that do not extend outside the confines of the star. In the second stage, the proton remains always inside the star interacting time after time until its kinetic energy becomes equal to the corresponding thermal energy with a temperature equal to that of the star's interior. To estimate the overall time scale, we evaluate the time scales of both stages separately. We follow closely the thermalization time scale derivation of~\cite{Kouvaris:2010jy}. Let's consider the first stage. The time between two collisions of the proton within the star takes place every $\Delta t$ given by half the period of the orbit of the proton as it travels in and out of the star divided by the ratio of the proton-dark matter cross section to the critical cross section.
	This critical cross section denotes the cross section value above which, on average, a proton crossing the star will scatter off at least a dark matter particle. It is given by $\sigma_{\text{crit}}=m R_{\rm DS}^2/M_{\rm DS}$ with $m$, $M_{\rm DS}$ and $R_{\rm DS}$ being the dark matter mass, the star's mass and star's radius respectively. This critical cross section corresponds (by construction) to a proton mean free path $\sim R_{\rm DS}$ i.e., as mentioned earlier on average there is going to be one scattering per crossing. Obviously for cross sections smaller than $\sigma_{\text{crit}}$ the probability of scattering is given by the ratio of the two.
	For simplicity we are going to assume that the proton orbit passes through the center of the star. Every time the proton scatters it loses energy of the order of $\sim 2 m/m_p$ of its kinetic energy (assuming that $m<<m_p$). After averaging over the point where the scattering takes place, the energy loss of the proton is given by
	\begin{equation}
		\Delta E= 2G M_{\rm DS} m\left (\frac{4}{3R_{\rm DS}}-\frac{1}{r} \right ),
	\end{equation}
	where $r$ is the size of the proton orbit. Dividing this energy change by $\Delta t$ and expressing $r$ in terms of energy we get a differential equation that dictates the evolution of the proton energy
	\begin{equation}
		\frac{dE}{dt}=-\frac{2\sqrt{2}m\sigma}{\pi G M_{\rm DS} m_p^{5/2}}\left (\frac{4}{3}E_*+E\right )|E|^{3/2},   
	\end{equation}
	where $E_*=GM_{\rm DS}m_p/R_{\rm DS}$ is the binding energy of the proton at the surface of the star. From this equation we can estimate the time scale for the first stage to be
	\begin{equation}
		t_1=\frac{\pi m_p R_{\rm DS}^{3/2} \sigma_{\text{crit}}}{2m\sqrt{2GM_{\rm DS}}\sigma}\int_{\epsilon_0}^1\frac{d\epsilon}{(4/3-\epsilon)\epsilon^{3/2}}\sim \frac{3 \pi m_pR_{\rm DS}^{3/2}\sigma_{\text{crit}}}{4 m \sqrt{2GM_{\rm DS}}\sigma}\sqrt{\frac {E_*}{|E_0|}},
	\end{equation}
	where $\epsilon_0$ is the ratio of $E_0$ (the initial energy of the proton) to  $E_*$. Numerically $\epsilon_0 \sim m/m_p$ is a small quantity and therefore
	\begin{equation}
		t_1\simeq 10^{-2}\text{s} \left(\frac{R_{\rm DS}}{1\ \text{km}} \right)^{7/2}\left (\frac{M_{\rm DS}}{M\odot} \right )^{-3/2}\left (\frac{\sigma}{10^{-50}\ \text{cm}^2} \right )^{-1}\left (\frac{m}{0.1\ \text{GeV}} \right )^{-1/2}.
	\end{equation}
	At the second stage the time between two successive collisions is $\Delta t=1/(n \sigma v)=m/(\rho \sigma v)$ where $n$ and $\rho$ are the number and mass density of dark matter particles in the star and $v=\sqrt{2 E/m_p}$.
	The rate of energy loss can be approximated by
	\begin{equation}
		\frac{d E}{dt}\simeq \frac{\Delta E}{\Delta t}\simeq -2\sqrt{2}\rho\sigma\left(\frac{E}{m_{\mathrm{p}}}\right)^{\frac32},
	\end{equation}
	where we have used that the energy loss in a collision is $\Delta E \simeq -2 m/m_{\rm p} E$. 
	The time scale for the second stage is then given by
	\begin{equation}
		t_{2}=\frac{m_{\mathrm{p}}}{\sqrt{2}\rho \sigma }\left(\sqrt{\frac{m_{\mathrm{p}}}{E_{\mathrm{final}}}}-\sqrt{\frac{m_{\mathrm{p}}}{E_{\mathrm{initial}}}}\right),
	\end{equation}
	which depends on the dark matter density, and is thus position dependent. Taking for simplicity a uniform density, $\rho=3M_{\rm DS}/(4\pi R_{\rm DS}^3)$ one obtains
	\begin{equation}
		t_2\approx 2 \times10^{-2}\ \text{s}
		\left(\frac{R_{\rm DS}}{1\ {\rm km}}\right)^{3}
		\left(\frac{M_{\rm DS}}{M_\odot}\right)^{-1}
		\left(\frac{\sigma}{10^{-50}\ {\rm cm}^2}\right)^{-1}
		\left(\frac{T}{10^{12}\ {\rm K}}\right)^{-1/2}.
	\end{equation}
	The total thermalization time scale is the sum of $t_1$ and $t_2$ which are of the same order of magnitude and in any case of interest much shorter than any other relevant process. Therefore, protons and subsequently electrons through Coulomb interactions once captured will very quickly thermalize and obtain the same temperature with the interior of the star.

	Once thermalized, the protons will concentrate within a radius $r_{\rm th}$ (the ``thermal radius") around the center of the dark star. The thermal radius can be estimated using the virial theorem
	\begin{equation}
		2\langle E_{\rm kin}\rangle=-\langle V_{\textrm{DS}} \rangle,
	\end{equation}
	where $\left<E_{k}\right>$ is the average total kinetic energy of the proton cloud and $\langle V_{\textrm{DS}}\rangle$ is the average total potential energy of the proton cloud in the gravitational potential of the dark star. The average total kinetic energy can be estimated from the equipartition theorem
	\begin{align}
		\langle E_{\rm kin} \rangle \approx \frac{3}{2}NT, 
	\end{align}
	with $N$ the number of protons within the thermal radius and $T$ the temperature of the proton gas (equal to the temperature of the dark star). To calculate the average potential energy, we assume for simplicity that the proton gas has a constant density. Then, considering that the thermal radius is expected to be much smaller than the radius of the dark star, so that the dark matter density is approximately constant within the thermal radius, the average total potential energy reads
	\begin{align}
		\langle V_{\textrm{DS}}\rangle\approx\frac{4\pi}{5}G\rho_{\mathrm{core}}m_{\mathrm{p}}r_{\mathrm{th}}^2,
	\end{align}
	where $G$ is Newton's constant, $m_p$ is the mass of the proton, and where we have used that the proton density is $\rho_{\mathrm{p}}=\frac{N m_{\mathrm{p}}}{\frac{4\pi}{3}r^3_{\mathrm{th}}}$. Applying the virial theorem, we obtain that the thermal radius of the proton cloud depends on the temperature at the dark star core through
	\begin{equation}\label{ThermalRadius}
		r_{\mathrm{th}}\approx\sqrt{\frac{15 T}{4\pi G \rho_{\mathrm{core}}m_{\mathrm{p}}}}.
	\end{equation} 
	Numerically, 
	\begin{align}
		r_{\mathrm{th}}\approx 0.4\, {\rm km}\Big(\frac{\rho_{\rm core}}{10^{18}\ {\rm gr}/{\rm cm^3}}\Big)^{-1/2} \left(\frac{T}{10^{12}\ K}\right)^{1/2}.
	\end{align}
	A comment is in order here. As mentioned earlier, although we study the accretion of protons into the dark star, electrons will be also accreted at an equal rate such that the dark star remains electrically neutral. From the dependence of $r_{\rm th}$ on the particle mass, one would naively expect that the thermal radius of the electrons should be $10^{1.5}$ times larger than that of protons. However, there should be no separation of charge because this induces large Coulomb fields with a high energy cost. Protons and electrons are tightly coupled and therefore they share the same thermal radius. From the virial theorem, protons contribute most of the mass and therefore the thermal radius is dominated by them and not the electrons. Therefore we expect our final formula to be true for the proton-electron plasma. We have additionally checked that for the parameters of interest used here, the electrons accreted into the star and confined within the thermal radius do not become degenerate and therefore no Fermi pressure is induced.  
	
	The hot and dense proton cloud will emit thermal radiation, with an intensity given by \citep{Ghisellini:2012hs}
	\begin{equation}\label{Intensity}
		I(\nu)=\frac{1}{4\pi^3}\frac{\epsilon^3}{\mathrm{e}^{\frac{\epsilon}{T}}-1}\left(1-\mathrm{e}^{-\tau(\epsilon)}\right),
	\end{equation}
	with $\epsilon$ the energy of the radiation and $\tau(\epsilon)$ the optical depth, which accounts for the possible absorption of the emitted photons when they propagate through the thermal sphere. The optical depth reads
	$\tau(\epsilon)\equiv r_{\rm th} \alpha_\epsilon$, with $\alpha_\epsilon$ the absorption coefficient, given by \cite{Ghisellini:2012hs}
	\begin{equation}\label{AbsorptionCoeff}
		\alpha_{\epsilon}=\frac{16\pi^2}{3}\left(\frac{2\pi}{3}\right)^{\frac12}\bar{g_{ff}}\frac{\alpha^3}{m_e^2}n_{e}n_{p}\left(\frac{m_e}{T}\right)^{1/2}\frac{1-\mathrm{e}^{-\frac{\epsilon}{T}}}{\epsilon^3},
	\end{equation}
	where $\alpha\approx1/137$ is the fine structure constant. Here, $n_{p}$ and $n_{e}$ are the number densities of  protons and electrons within the thermal sphere; the former is determined by the capture rate via the scattering dark matter-proton, and the latter by the Coulomb capture by protons (electric neutrality ensures  $n_e= n_p$). The factor $\bar{g_{ff}}$ is the ``Gaunt'' factor, which lies within $1.1$ and $1.5$. For our purposes, we have taken $\bar{g_{ff}}=1.3$. 
	Assuming that at the time of star formation there were no protons inside the dark star, one can approximate $n_p\simeq C t/(4/3 \pi r_{\rm th}^3)$, with $C$ the capture rate, which is given in Eq.~(\ref{eq:capture_num}). 
	
	Finally, the photon luminosity of a dark star with core temperature $T$ can be calculated by integrating the intensity over all energies, and multiplying by the area of the thermal sphere:
	\begin{equation}\label{eq:LightLuminosity}
		L_{\gamma}=\left(4\pi r_{\mathrm{th}}^2\right)\int_{0}^{\infty}{I(\epsilon)d\epsilon}\;.
	\end{equation}
	When the gas of electrons and protons within the thermal sphere is optically thick, so that $\tau(\epsilon)\gg 1$ for all energies, the emitted radiation follows a black body spectrum, and has a total intensity given by the Stefan-Boltzmann law. Therefore, the total luminosity simply reads
	\begin{equation}\label{eq:LightLuminosity2}
		L_{\gamma}=\left(4\pi r_{\mathrm{th}}^2\right)\sigma_{\rm SB} T^4\;,
	\end{equation}
	with $\sigma_{\rm SB}$ the Stefan-Boltzmann constant. Note that since $r_{\rm th}\sim T^{1/2}$, one finds a photon luminosity with a strong temperature dependence $\sim T^5$. In contrast, when the gas of protons and electrons is optically thin, the bremsstrahlung photons exit the thermal sphere without  any interaction, thus preserving their initial bremsstrahlung spectrum. In this case, the intensity is suppressed by the small factor $(1-e^{-\tau(\epsilon)})$, which results in a much lower luminosity than in the optically thick regime. 
	
	Furthermore, in many models the dark star would also emit ``dark radiation". In fact, the emission of dark photons could be pivotal for the formation of the dark star. In order for the asymmetric dark matter particles to collapse and form the assumed compact dark stars, an efficient mechanism of energy evacuation is needed. If energy cannot be evacuated effectively, dark matter will form a halo and not a compact object. Dark photon emission via e.g. dark bremsstrahlung radiation can provide such an effective mechanism that can facilitate the dark star formation~\cite{Chang:2018bgx}.
	
	For a dark photon with mass $m_{\gamma'}$, and assuming for simplicity that the dark star is optically thick to the dark radiation, one finds a dark photon luminosity given by
	\begin{equation}\label{eq:DPLuminosity}
		L_{\gamma'}=(4\pi R_{\mathrm{DS}}^2)\, \sigma_{\mathrm{SB}}T^4\mathrm{e}^{-\frac{m_{\gamma'}}{T}},
	\end{equation}
	where we have included an exponential suppression of the luminosity when the temperature of the dark star falls below the dark matter mass. Note that the dark photon radiation is emitted from the surface of the dark star, while the visible photon radiation is emitted from the surface of the thermal sphere.  
	
	Clearly, the photon  and dark photon luminosities are not constant in time, since the emission of photons and dark photons themselves cools down the dark star. In the next section we will analyze in detail the temperature evolution of the dark star, which in turn determines the time evolution of the photon and dark photon luminosities. 
	
	\section{Temperature evolution of the dark star}\label{sec:Thermal_evolution}
	
	The internal energy of the dark star decreases with time due to the emission of visible and dark photons:
	\begin{align}
		\frac{dU}{dt}=-(L_\gamma+L_{\gamma'})\;,
	\end{align}
	with the visible and dark photon luminosities given in Eqs. (\ref{eq:LightLuminosity}) and (\ref{eq:DPLuminosity}).
	In turn, the internal energy depends on the dark star temperature. Therefore, one can cast the temperature evolution of the dark star as 
	\begin{equation}\label{TemperatureEvolutionEquation}
		\frac{dT}{dt}=-\frac{L_{\gamma}+L_{\gamma'}}{C_{\mathrm{v}}},
	\end{equation}
	where $C_V\equiv (dU/dT)_V$ is the heat capacity of the dark star (assuming constant size).
	
\begin{figure}[t! ]%
		\centering
		\includegraphics[width=.48\textwidth]{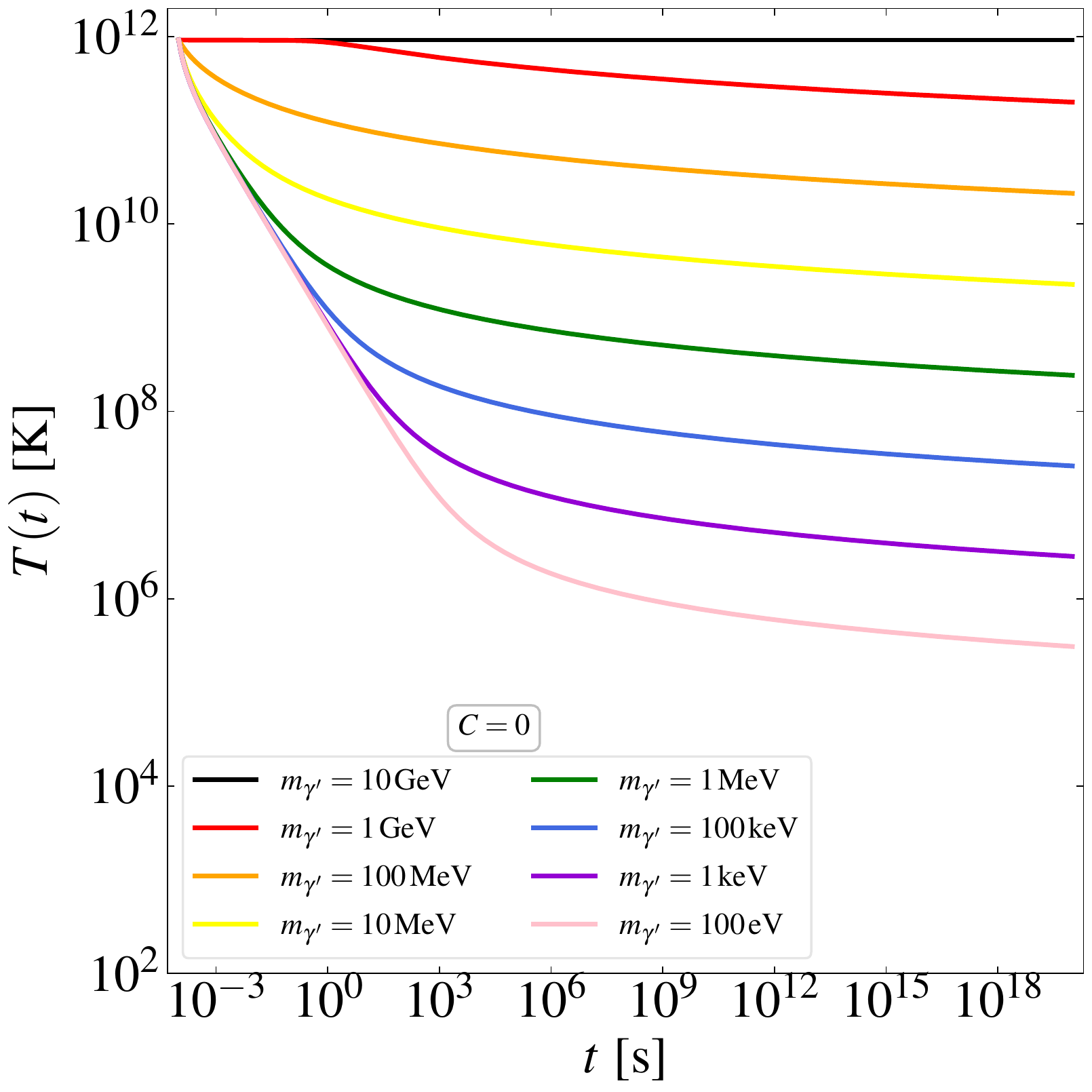}
		\includegraphics[width=.48\textwidth]{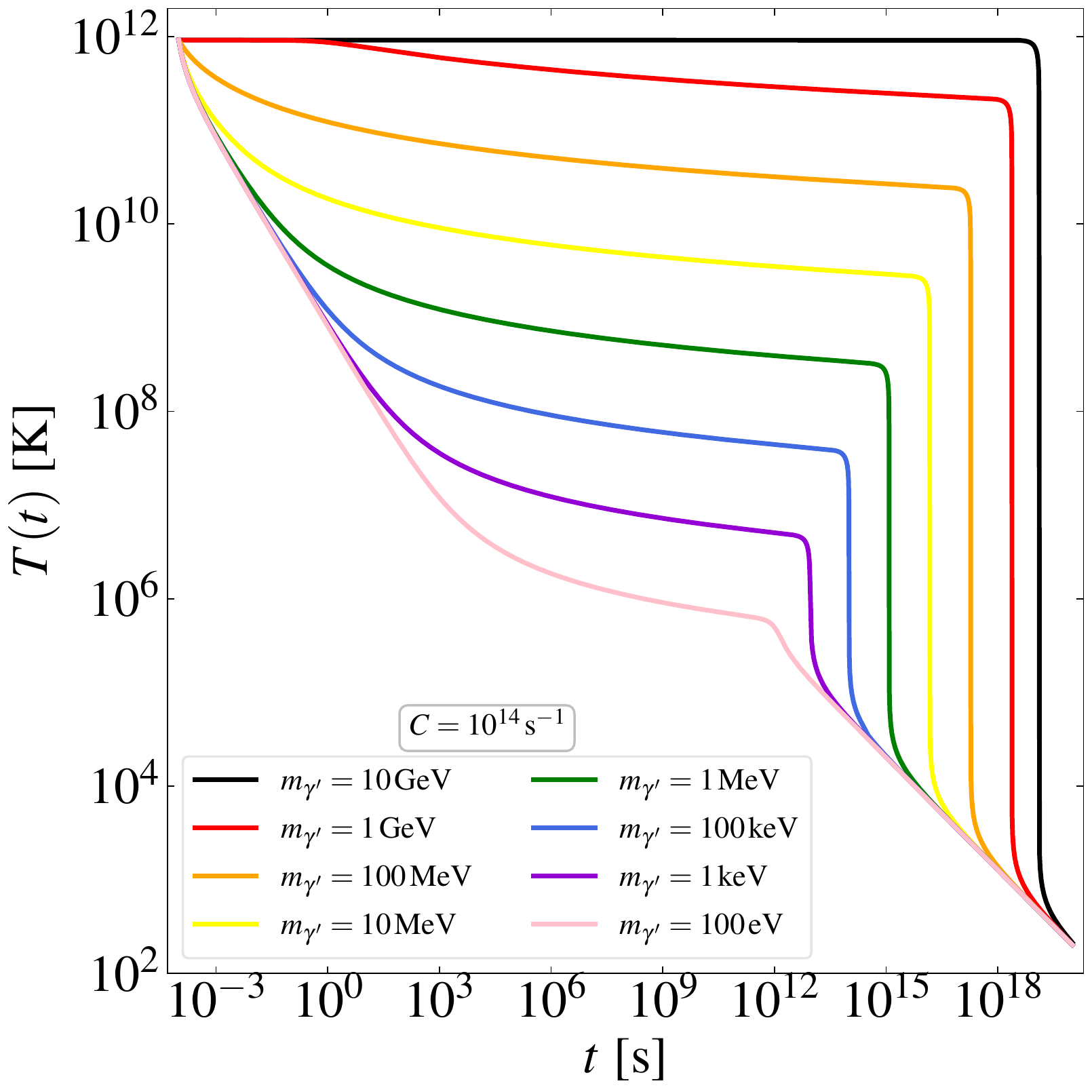}\\
		\includegraphics[width=.48\textwidth]{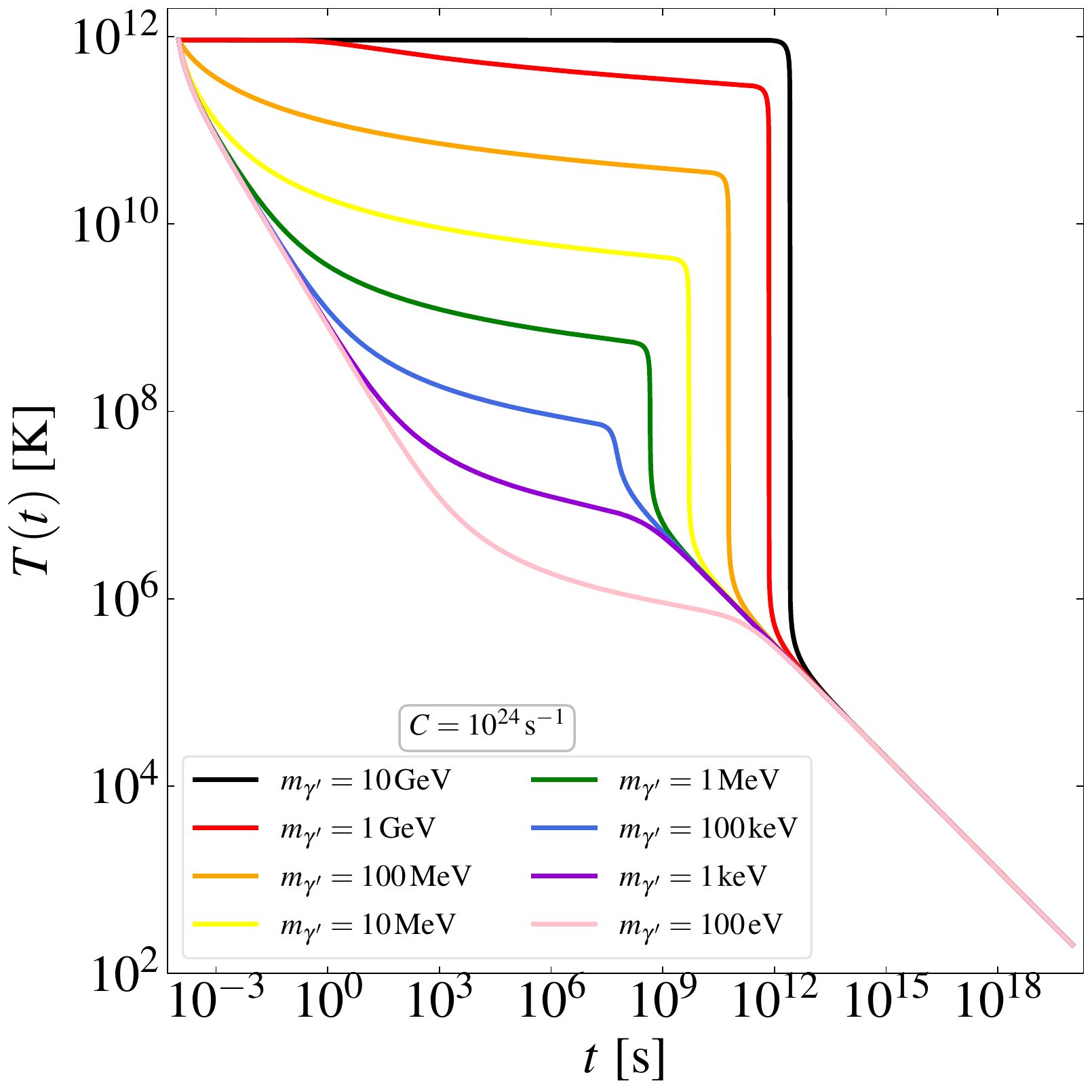} 
		\includegraphics[width=.48\textwidth]{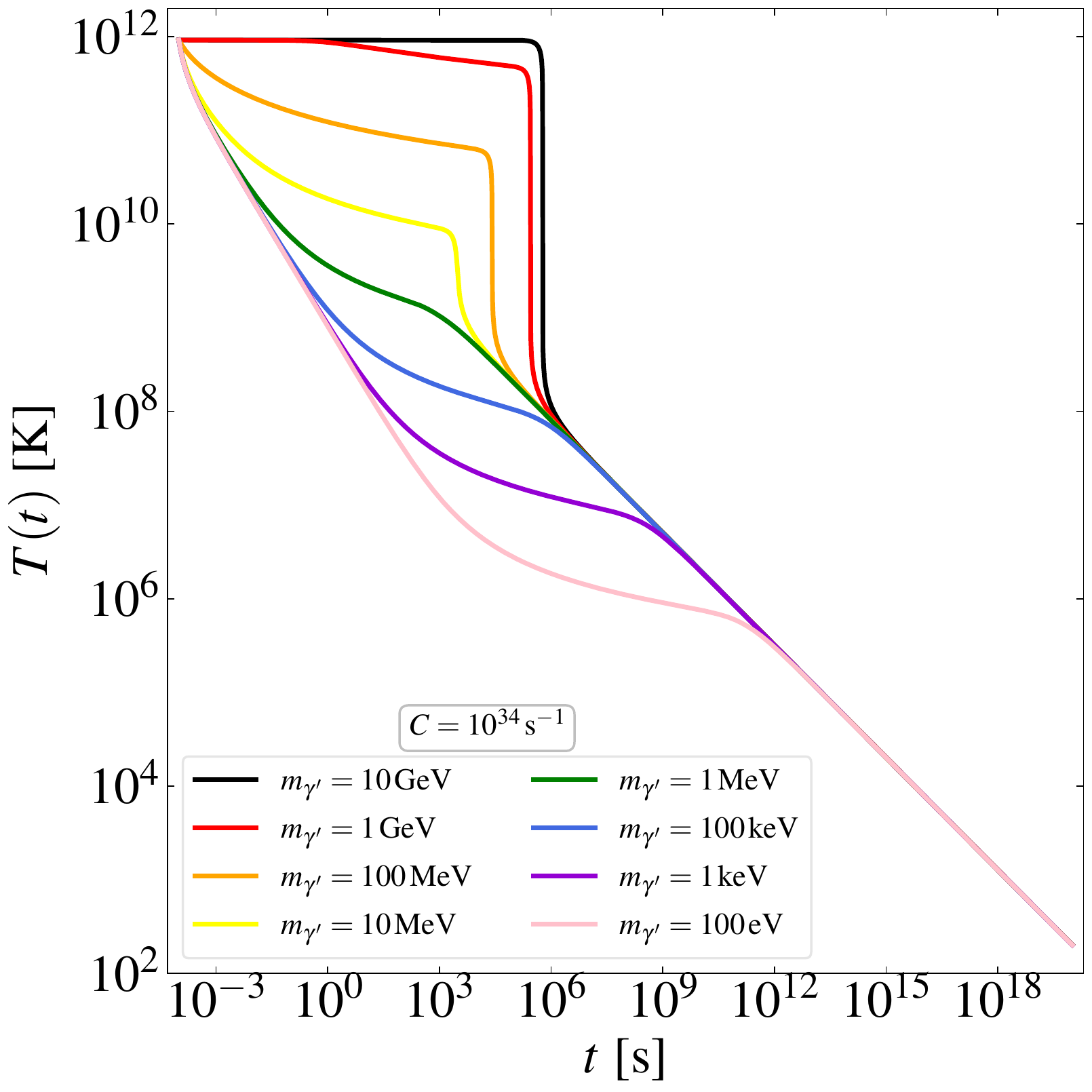}
		\caption{Temperature evolution of a compact dark star as a function of time, due to the emission of dark photons with mass $m_{\gamma'}$, and due to the emission of photons from a gas of captured protons (and electrons), when the capture rate is $C=0$ (top left), $10^{14}\,{\rm s}^{-1}$ (top right), $10^{24}\,{\rm s}^{-1}$ (bottom left) and $10^{34}\,{\rm s}^{-1}$ (bottom right).}
		\label{fig:Temperature_vs_time}%
	\end{figure}

\begin{figure}[t! ]%
		\centering
		\includegraphics[width=.48\textwidth]{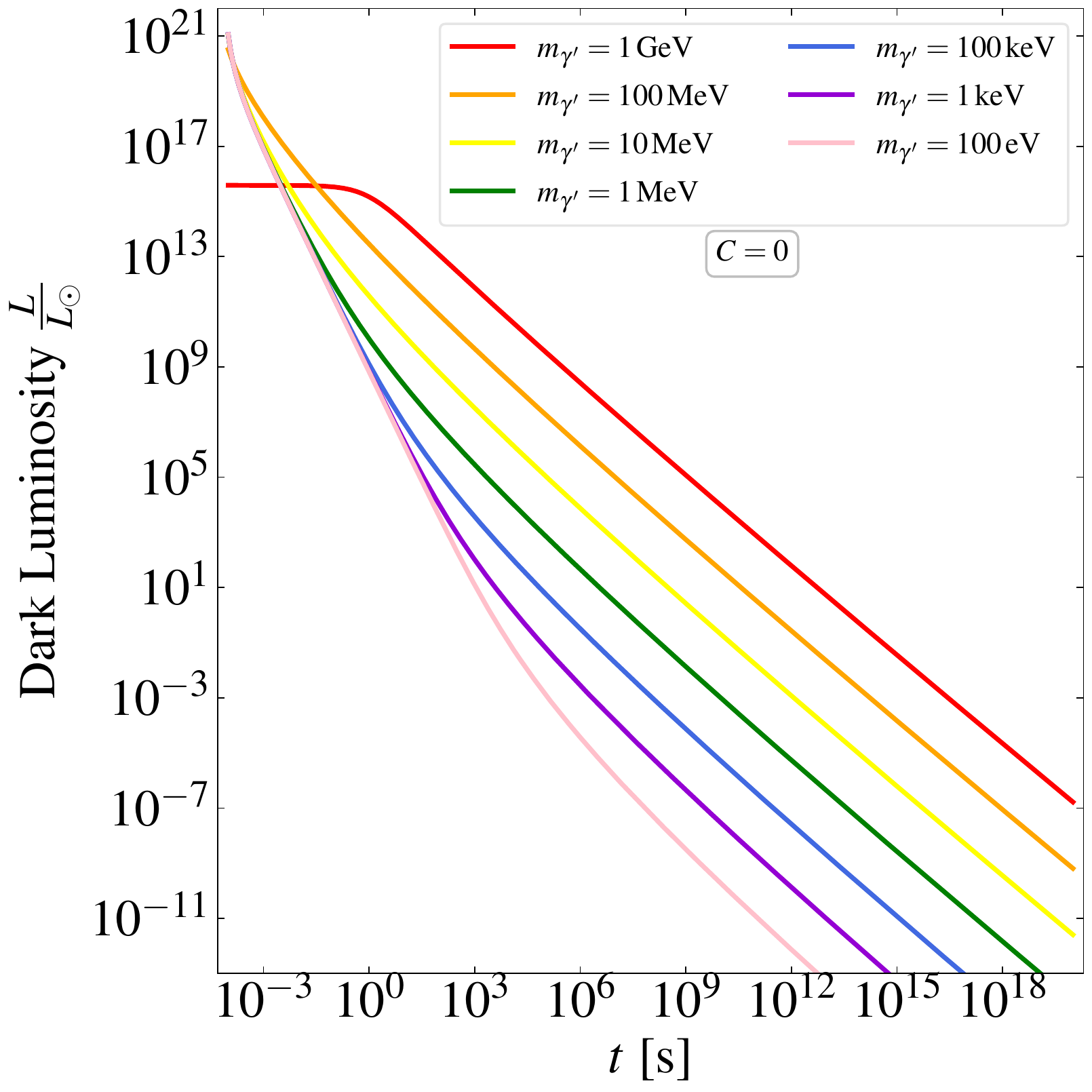}
		\includegraphics[width=.48\textwidth]{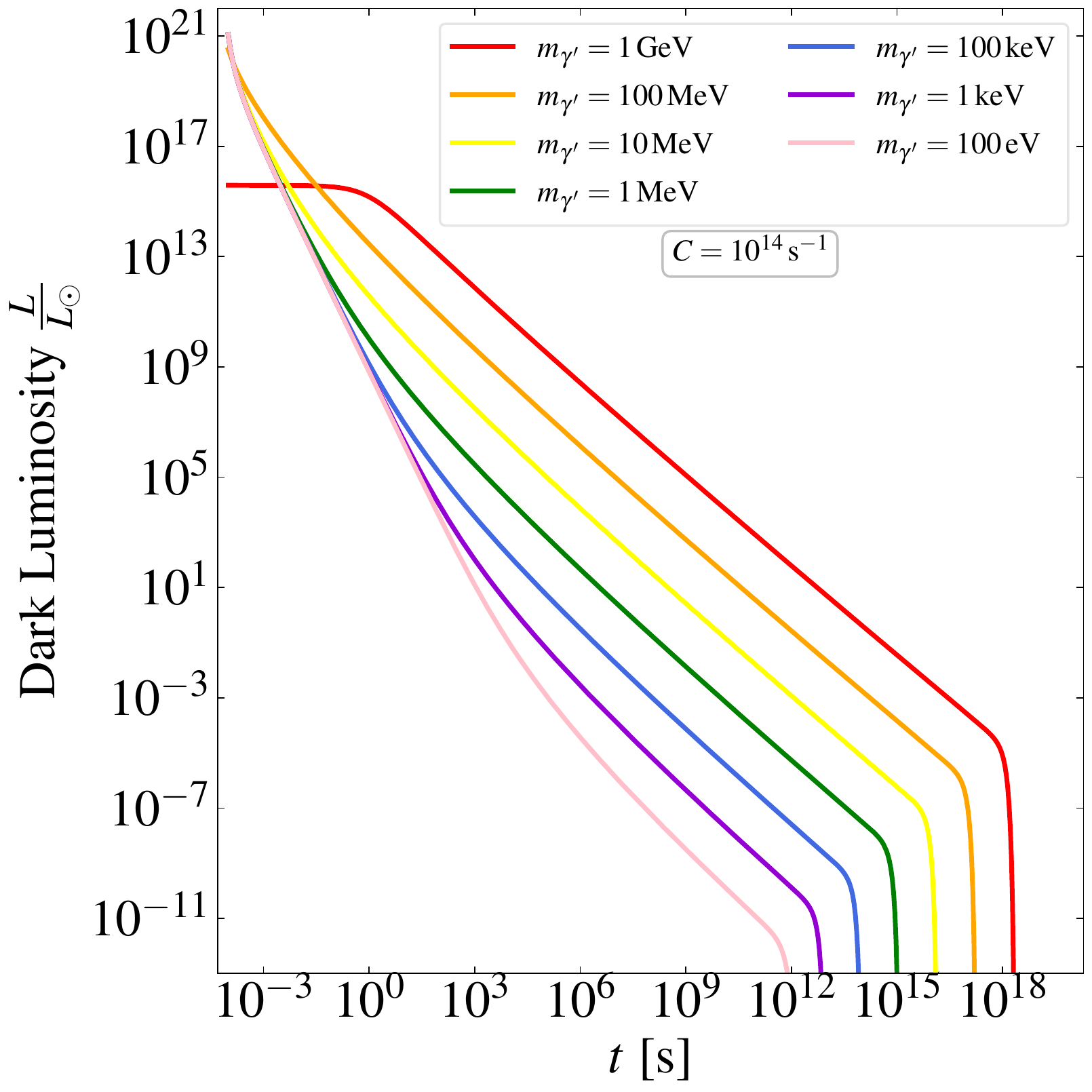}\\
		\includegraphics[width=.48\textwidth]{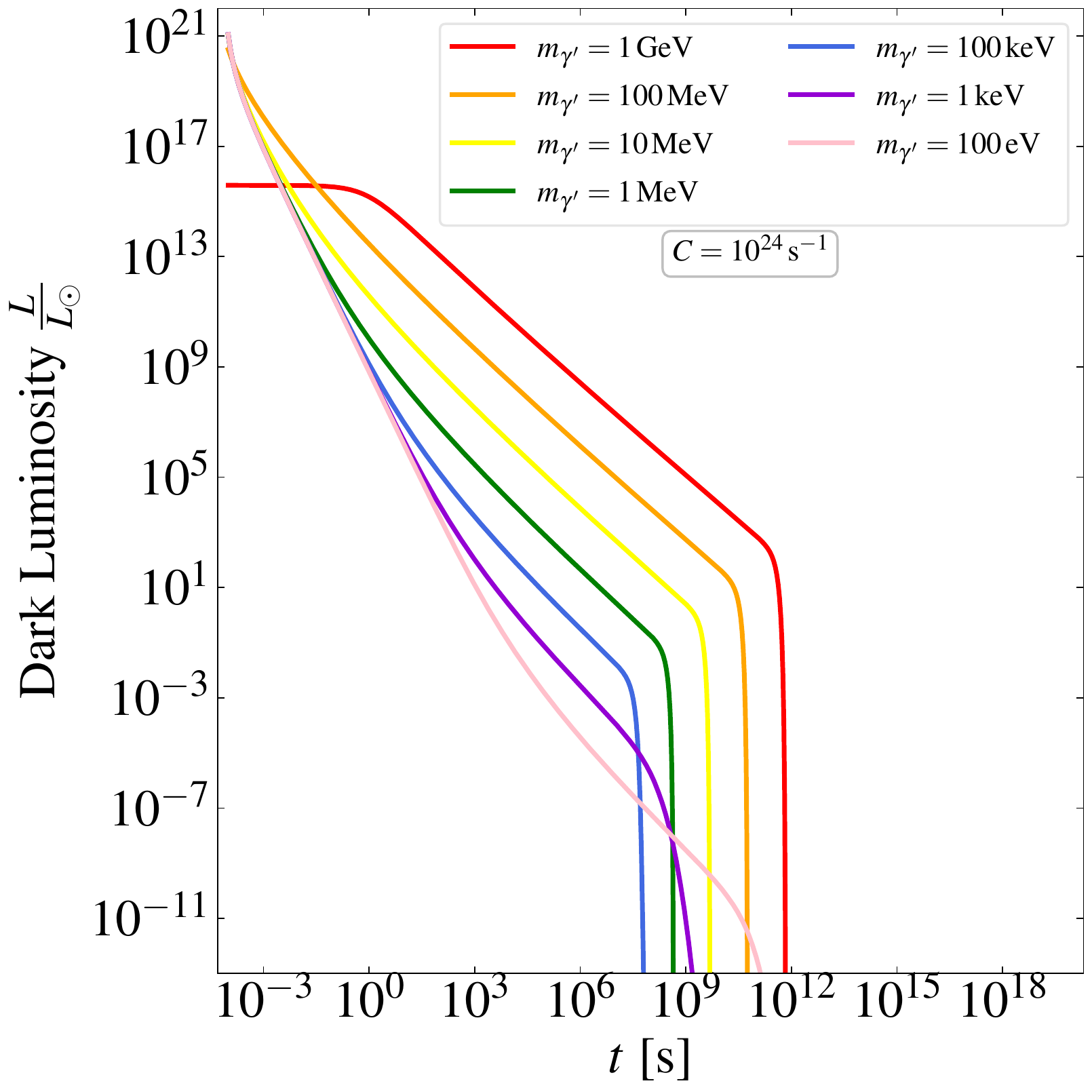}
		\includegraphics[width=.48\textwidth]{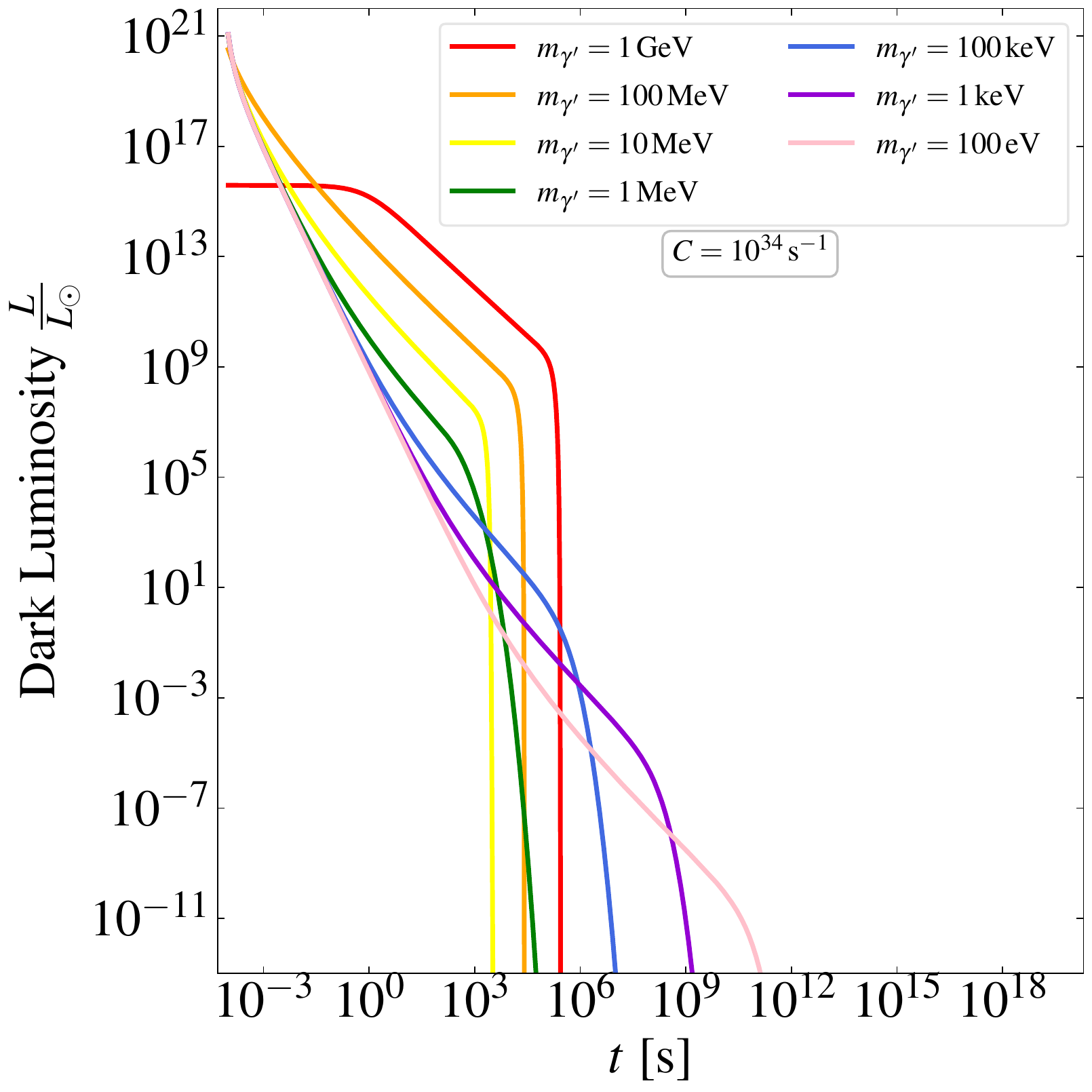}\caption{Same as Fig.~\ref{fig:Temperature_vs_time}, but showing the luminosity evolution in dark photons (normalized to the solar luminosity).}
		\label{fig:dark_photon_luminosity}%
	\end{figure}

\begin{figure}[t! ]%
		\centering
		\includegraphics[width=.48\textwidth]{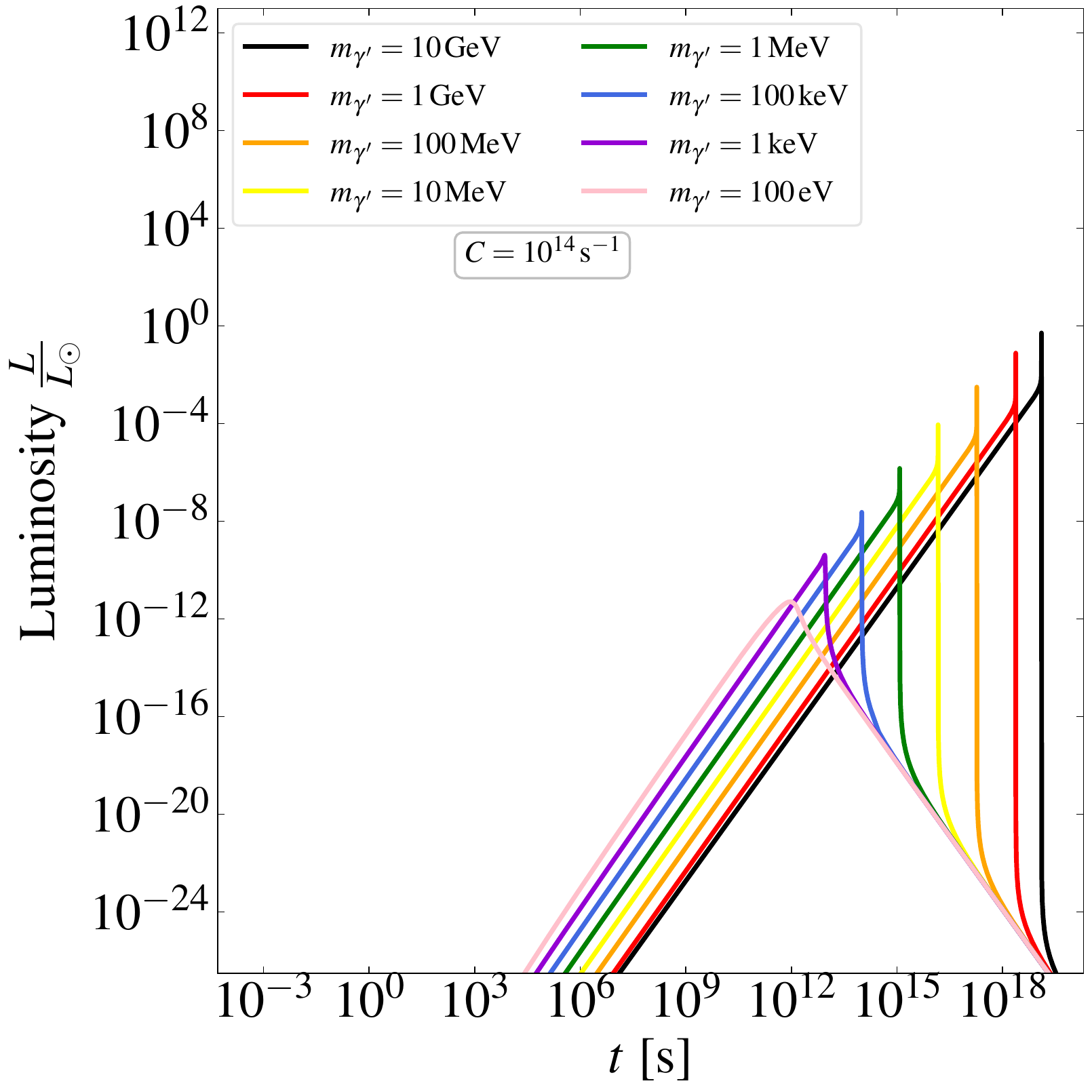}\\
		\includegraphics[width=.48\textwidth]{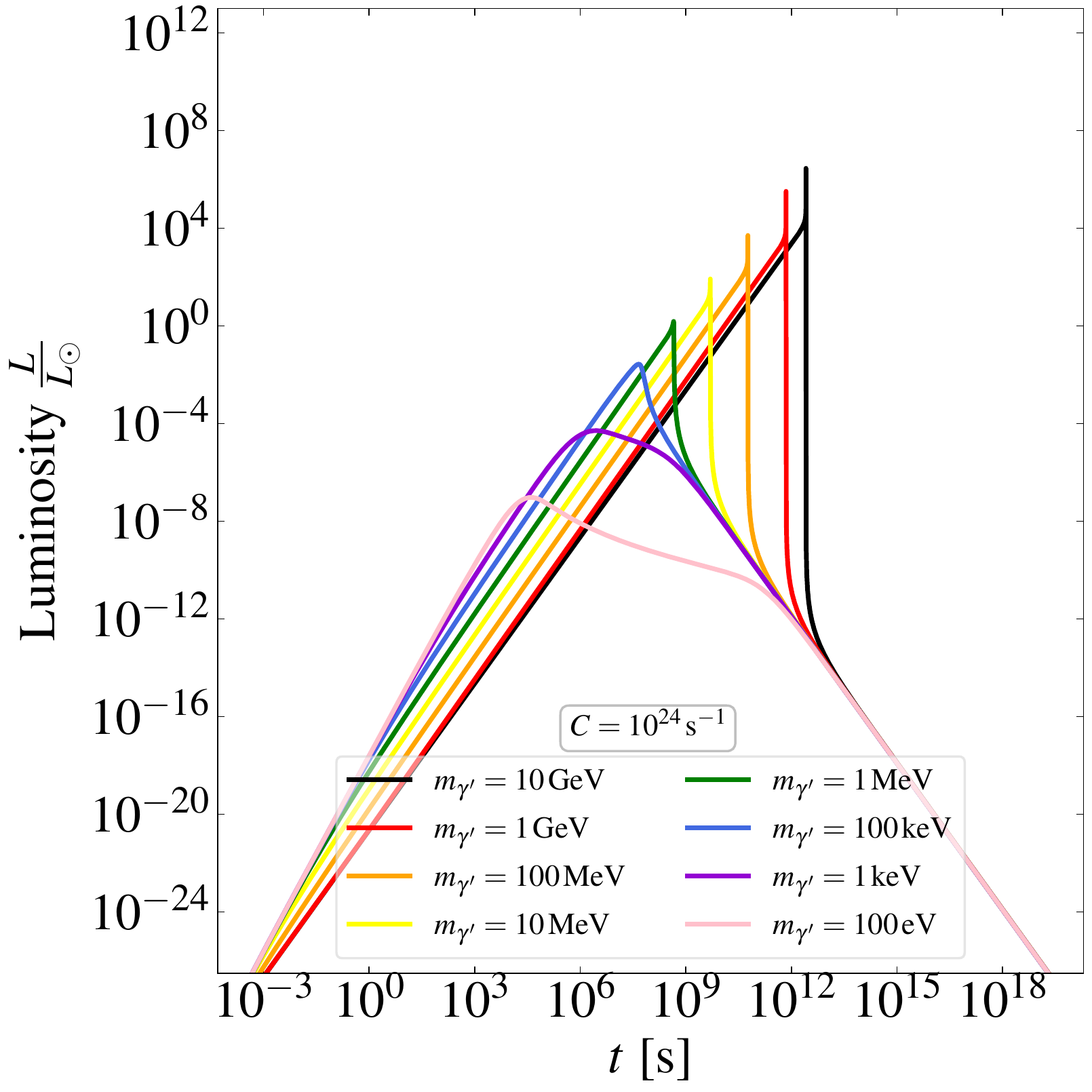} 
		\includegraphics[width=.48\textwidth]{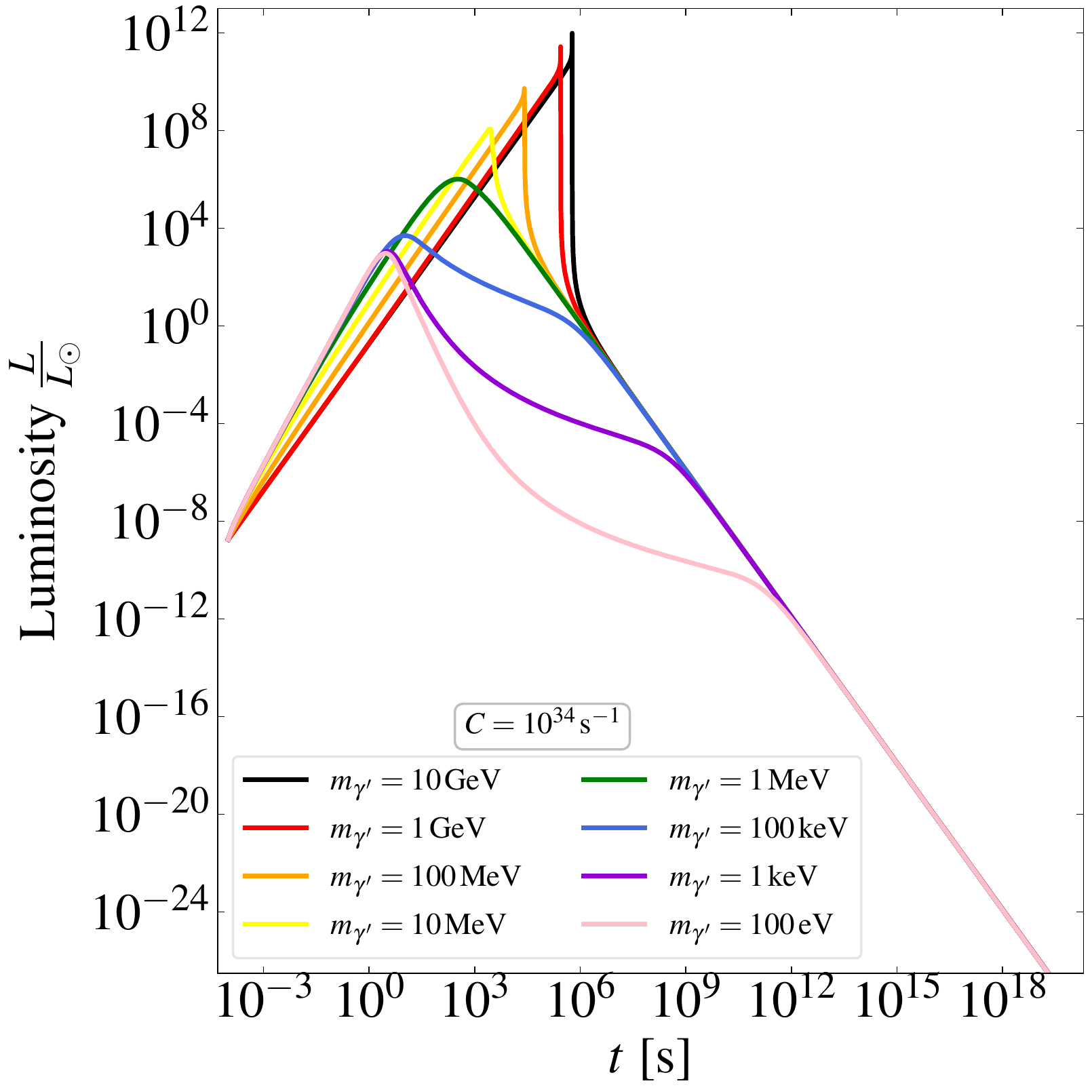}
		\caption{Same as Fig.~\ref{fig:Temperature_vs_time}, but showing the luminosity evolution in photons (normalized to the solar luminosity). The top panels do not show the case $C=0$, since the photon luminosity is zero at all times.}
		\label{fig:photon_luminosity}%
	\end{figure}

	The calculation of the heat capacity of a self gravitating system of interacting bosons is a complicated problem, well beyond the scope of this work. Therefore, for the calculation of the heat capacity, we crudely approximate the dark star as a gas of free bosons. The critical temperature of such a gas is
	\begin{equation}\label{CriticalTemperature}
		T_{\mathrm{c}}=\frac{2\pi}{m}\left(\frac{n}{\xi(\frac32)}\right)^{2/3},
	\end{equation}
	with $n$ the dark matter number density. To estimate the critical temperature, we approximate $n=\rho_{\rm core}/m$, with $\rho_{\rm core}$ the core density of the dark star and $m$ the boson mass.~\footnote{Strictly, the critical temperature is position dependent, due to the radially varying density.} Using the approximate formulas from (\ref{eq:Unit Conversion}) one finds
	
	\begin{equation}
		T_{\mathrm{c}}=(10^{14}\ \mathrm{K})\,\left(\frac{m}{1\ \mathrm{GeV}}\right)^{-\frac53}\left(\frac{\rho_{\rm core}}{10^{18}\ \mathrm{gr}/\mathrm{cm}^3}\right)^{\frac23}.
	\end{equation}
	Below this temperature the boson gas will form a Bose-Einstein condensate. We assume in what follows $T<T_{\mathrm{c}}$ at the formation of the dark star. Therefore, over the whole time evolution the heat capacity takes the well known value
	\begin{equation}\label{eq:HeatCapacity}
		C_{\mathrm{v}}=\frac{15}{4}N\frac{\xi(\frac52)}{\xi(\frac32)}\left(\frac{T}{T_{\mathrm{c}}}\right)^{\frac32}.
	\end{equation}
	The temperature evolution of the dark star can now be calculated from solving numerically Eq.~(\ref{TemperatureEvolutionEquation}).  The result is shown in Fig.~\ref{fig:Temperature_vs_time} for capture rate $C=0$, $10^{14}$, $10^{24}$ and $10^{34}\,{\rm s}^{-1}$. From the temperature, one can readily calculate the luminosity in dark photons and visible photons; the result is shown in Fig.~\ref{fig:dark_photon_luminosity} and Fig.~\ref{fig:photon_luminosity}, respectively (for $C=0$ the visible photon luminosity is zero at all times).
	
	If the capture rate is zero, the dark star can cool down only via the emission of dark photons. Concretely, for $T\gg m_{\gamma'}$ the temperature evolution is $dT/dt =  - a T^{5/2}$, with $a$ calculable from Eqs. (\ref{eq:DPLuminosity}, \ref{TemperatureEvolutionEquation}, \ref{eq:HeatCapacity}). The solution to this differential equation reads
	\begin{align}
		T(t)  =\frac{T_0}{{(1+\frac{3}{2} \,a \,t T_0^{3/2})}^{2/3}},
	\end{align}
	namely $T\sim T_0$ for $t\ll 2/(3\, a\, T_0^{3/2})$ and 
	$T\sim t^{-2/3}$ for $t\gg 2/(3\, a T_0^{3/2})$. The latter relation holds as long as $T\ll m_{\gamma'}$. When the temperature becomes comparable to the dark photon mass, the time dependence of the temperature reads instead $dT/dt= -a T^{5/2} \exp(-m_{\gamma'}/T)$, where the Boltzmann suppression factor slows down the cooling rate. Eventually, for $T\ll m_{\gamma'}$, $dT/dt\simeq 0$ and the temperature stays constant. This behavior is shown in Fig.~\ref{fig:Temperature_vs_time}, top-left panel. The dark photon luminosity depends at all times on $T^4\,\exp(-m_{\gamma'}/T)$, inheriting the behavior described above for the temperature, as seen in  Fig.~\ref{fig:dark_photon_luminosity}, top-left panel.
	
	If the capture rate is non-zero, the dark star also cools down via the photon emission from the proton (and electron) gas. At early times, the number of protons captured is small, and the photon emission is very small (see Fig.~\ref{fig:photon_luminosity}, top panel), while the dark photon emission is sizable (see Fig.~\ref{fig:dark_photon_luminosity}, top-right panel). As a result, the dark star cools down dominantly by the dark photon emission. As time evolves, the number of protons (and electrons) increases. However, the medium is still optically thin and the power emitted in visible photons is comparatively small compared to the power emitted in dark photons. As the temperature keeps reducing, it eventually might drop below the dark photon mass. In that case the dark photon emission becomes suppressed. Photon emission becomes dominant and as long as the thermal sphere is optically thin to photons, the luminosity which is proportional to $n_p^2 T^2$ scales as $t^2/T$ since the number of protons increases linearly with the time and the thermal radius (thermal volume) as $T^{1/2}$ ($T^{3/2}$). As the temperature drops, the thermal radius decreases. In turn, the shrinking thermal radius  causes  further increase of the luminosity and even faster energy loss and temperature fall (see Fig.~\ref{fig:Temperature_vs_time}) culminating to an outburst of photons seen in   Fig.~\ref{fig:photon_luminosity}.
	The outburst ends once the thermal sphere becomes opaque to photons due to the continuously decreasing  photon mean free path which in turn happens because of the decreasing thermal radius. Once the thermal sphere becomes opaque, it emits blackbody photons only from its surface with a total luminosity scaling as $T^5$ (resulting from the usual $T^4$ blackbody contribution and the $T$ dependence of the surface factor $r_{\rm th}^2$). At this point, reducing the temperature results also in a corresponding reduction of the luminosity ending the outburst.
	
	After the end of the outburst, the dark star emits dominantly visible photons, with luminosity (as mentioned above) $L_\gamma\sim T^5$, and cools down  at a rate $dT/dt\sim - T^{7/2}$, resulting in $T\sim t^{-2/5}$. In this regime, the photon luminosity scales as $\sim t^{-2}$. Notice that the temperature evolution of the dark star at late times is independent of the dark photon mass. This is due to the fact that once the proton gas becomes optically thick, it is characterized only by its temperature, and all previous history of the dark star (in particular how the proton gas becomes a black body) becomes irrelevant. 
	
	The bottom panels of Figs.~\ref{fig:Temperature_vs_time},  \ref{fig:dark_photon_luminosity} and \ref{fig:photon_luminosity} show respectively the temperature, dark photon luminosity and visible photon luminosity for $C=10^{24}\,{\rm s}^{-1}$ and  $C=10^{34}\,{\rm s}^{-1}$. The results are qualitatively similar as for $C=10^{14}\,{\rm s}^{-1}$.  However, due to the larger capture rate, the proton gas becomes optically thick earlier, and correspondingly the outburst occurs earlier. The outburst also occurs when the dark star is hotter, therefore the maximum luminosity of the dark star increases. Notice that at very late times, the dark star temperature and the photon luminosity are the same as for $C=10^{14}\,{\rm s}^{-1}$, due again to the fact that the dark star behaves as a black body, characterized only by its temperature (and therefore dependent only on the initial temperature of the dark star). 
	
	\section{Signals of compact dark stars}\label{sec:Signals}
	
\begin{figure}[t!]%
		\centering
		\includegraphics[width=.48\textwidth]{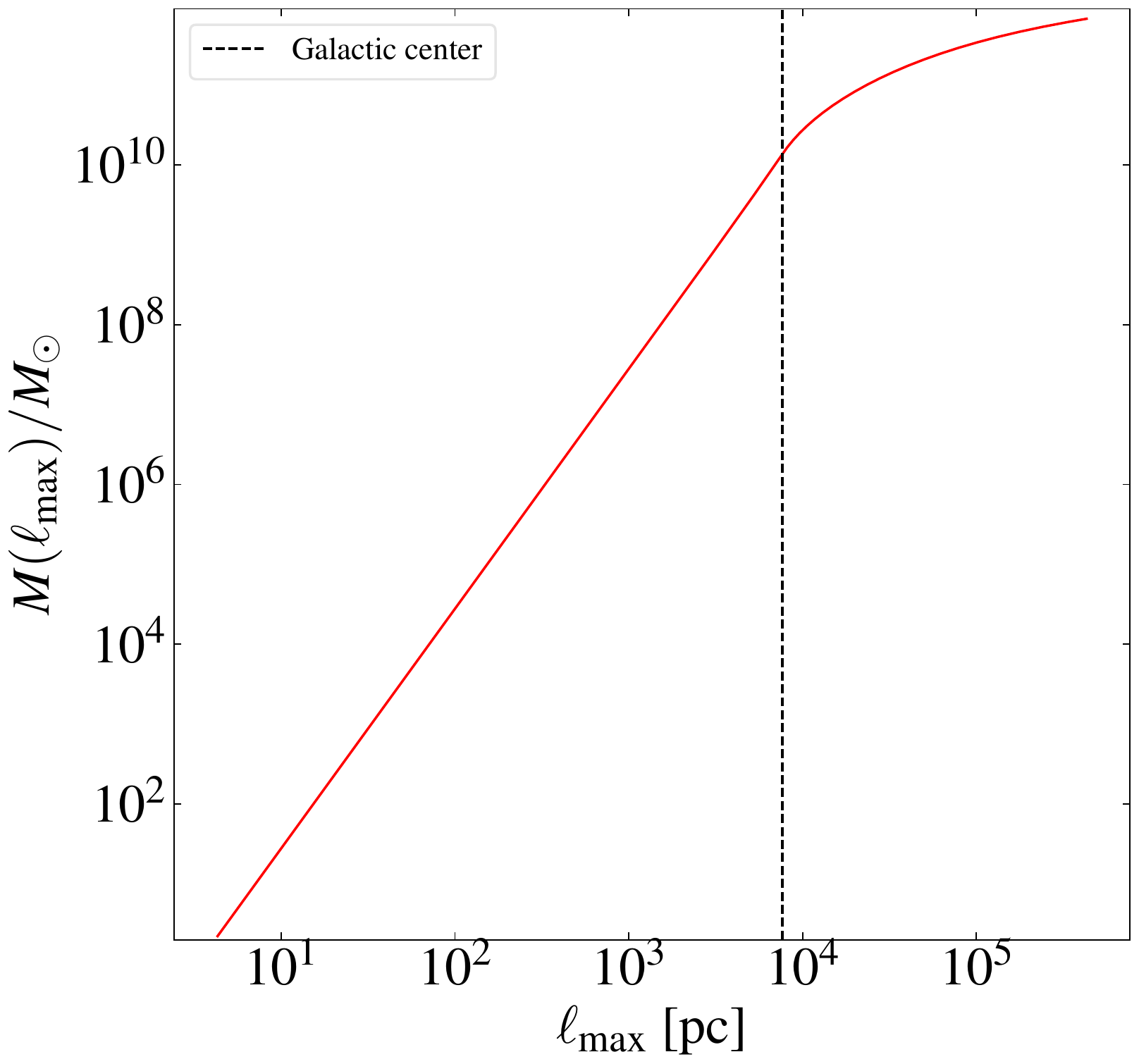}
		\caption{Enclosed dark matter mass (in units of the solar mass) within a sphere of radius $\ell_{\rm max}$ centered at the Earth.}
		\label{fig:EnclosedMassSphere}%
	\end{figure}

	The photons emitted by the dark star could be detected at the Earth as a point source in $\gamma$-rays.  To assess the prospects to observe the photon emission from dark stars in an instrument with sensitivity  to point sources $S_{\rm point}$, let us assume for simplicity that all dark stars evolve in the same way, {\it i.e.} their initial temperature, mass and radius are the same for all of them, as well as the proton capture rate. The maximum distance to the Earth of a dark star with photon luminosity $L_\gamma$ that can be detected with an instrument with point-source sensitivity $S_{\rm point}$ is given by
	\begin{align}
		{\ell}_{\rm max}=\left(\frac{L_\gamma}{4\pi S_{\rm point}}\right)^{1/2}.
		\label{eq:max_distance}
	\end{align}
	The number of dark stars within that distance is estimated from
	\begin{align}
		{\cal N}_{\rm DS}={\cal F}_{\rm DS} \frac{M(\ell_{\rm max})}{M_{\rm DS}},	
		\label{eq:number_DS_delta}
	\end{align}
	where $M(\ell_{\rm max})$ is the total dark matter mass enclosed in a sphere centered at the Earth of radius $\ell_{\rm max}$, and ${\cal F}_{\rm DS}$ is the fraction of the dark matter in the form of dark stars. We consider for concreteness that the dark matter in our Galaxy is distributed following a Navarro-Frenk-White density profile
	\begin{equation}
		\rho(r)=\displaystyle{\frac{\rho_0}{\frac{r}{r_s}\left(1+\frac{r}{r_s}\right)^2}},
	\end{equation}
	with $\rho_0=1.06\times10^{-2}\, M_\odot{\rm pc}^{-3}$ and $r_s=12.5\,\mathrm{kpc}$  \citep{Sofue:2011kw}. The mass enclosed within a distance $\ell_{\rm max}$ is
	\begin{align}
		M(\ell_{\rm max})=2\pi\int_{-1}^1 d\cos\theta \int_0^{\ell_{\rm max}} d\ell\, \ell^2 \rho\big(r(\ell,\cos\theta)\big),
	\end{align}
	with $r(\ell, \cos\theta)=(d_\odot^2+\ell^2-2 \ell\,d_\odot\,\cos\theta)^{1/2}$ and $d_\odot\simeq 8.5$ kpc the distance between the Earth and the Galactic Center. Explicitly
	\begin{equation}\label{eq:EnclosedMass}
		M(\ell_{\max})\simeq(3.3\times10^{11}M_\odot)
		\begin{cases}
			\displaystyle{\log{\left(\frac{\frac{41}{25}+\frac{\ell_{\rm max}}{r_s}}{\frac{41}{25}-\frac{\ell_{\max}}{r_s}}\right)}-\frac{50}{41}\frac{\ell_{\rm max}}{r_s}} & \text{if}~\ell_{\rm max}<d_\odot,\\
			\displaystyle{\log{\left(\frac{41}{25}+\frac{\ell_{\rm max}}{r_s}\right)}-\frac{9}{41}\log{\left(\frac{9}{25}+\frac{\ell_{\rm max}}{r_s}\right)}-\frac{32}{41}}&\text{if}~\ell_{\rm max}\geq d_\odot,
		\end{cases}
	\end{equation}
	which is shown in Fig.~\ref{fig:EnclosedMassSphere}. Taking ${\cal F}_{\rm DS}=10^{-2}$, compatible with current MACHO searches, and $S_{\rm point}=10^{-11}\,{\rm erg}\,{\rm cm}^{-2}\,{\rm s}^{-1}$, which is the typical point source sensitivity of current $\gamma$-ray telescopes, one obtains $\sim 1$ point source for a luminosity today $L_\gamma=10^{-4} L_\odot$ or $\sim 10^6$ point sources for $L_\gamma= L_\odot$. 
	Comparing with Fig.~\ref{fig:photon_luminosity} one finds several scenarios leading to these luminosities. For instance, a dark star would have today a luminosity $L_\gamma=10^{-4} L_\odot$ if the dark photon mass is  $m_{\gamma'}=10$ GeV, the capture rate is $C=10^{14}\,{\rm s}^{-1}$ and it was formed $\sim 10^{17}$ s ago (approximately at the same epoch as our Sun) and a luminosity $L_\gamma= L_\odot$ if the dark photon mass is  $m_{\gamma'}=100$ MeV, the capture rate is $C=10^{24}\,{\rm s}^{-1}$ and it was formed $\sim 10^{9}$ s ago. If all dark stars formed at the same epoch, one would expect ${\cal O}(1)$ and ${\cal O}(10^6)$ events respectively. 
	Dark stars could form continuously in our Galaxy, with formation rate $g(t)$. The luminosity today of a dark star formed at the time $t$ is $L_{\gamma}(t_0-t)$, with $t_0$ the age of the Universe. Therefore, an instrument with point source sensitivity $S_{\rm point}$ can detect a dark star formed at the time $t$ if it lies within a distance
	\begin{align}
		{\ell}_{\rm max}(t)=\left(\frac{L_\gamma(t_0-t)}{4\pi S_{\rm point}}\right)^{1/2}.
	\end{align}
	From the number of dark stars within that distance, given by Eq.~(\ref{eq:number_DS_delta}), and the formation rate, one obtains that the expected number of point sources reads
	\begin{equation}
		N_{\mathrm{today}}=\int_{0}^{t_0}dt\,g(t)\,\mathcal{F}_{\chi}\frac{M(\ell_{\rm max}(t))}{M_{\mathrm{DS}}},
	\end{equation}
	with $M(\ell_{\rm max}(t))$ given in Eq.~(\ref{eq:EnclosedMass}). If all dark stars form at the time $t=t_{\rm form}$, the previous formula reduces to Eq.~(\ref{eq:number_DS_delta}), with the luminosity in Eq.~(\ref{eq:max_distance}) evaluated at the time $t_0-t_{\rm form}$, {\it i.e.} the age of the dark star. For a general function $g(t)$ the solution is more involved and depends crucially on the model. As an example, if one assumes a constant rate of dark star formation, then dark star solutions which reach the outburst in a fast time scale (such as the solutions from the lower left panel in Fig. \ref{fig:photon_luminosity}) will have better prospects of detection compared to the case of a unique genesis, as discussed previously.
	We note that in our estimation we have included only the photons directly emitted by the hot gas of protons and electrons. In the presence of kinetic mixing, a fraction of the dark photons emitted would convert into ordinary photons, thus increasing the total photon flux from compact dark stars. This contribution could be especially important for dark stars formed in recent times, since right after the formation of the dark star the dark photon luminosity is many orders of magnitude larger than the photon luminosity. 
	
	\section{Dark matter annihilation or decay inside the compact dark star}\label{sec:Annihilation_Decay}
	
	So far we have assumed the existence of a ``dark matter number" (akin to the electric charge or the baryon number) that prevents the decay or the annihilation of dark matter particles into Standard Model particles. Observations show that the rate for the dark matter number violating processes must be sufficiently small, so that large amounts of dark matter are still present in galaxies and in the Universe at large. However, observations do not require the absolute conservation of the dark matter number. The high dark matter density of dark stars, together with the possible proximity of  such an object to the Earth, offers a unique environment to probe the conservation of the dark matter number.
	
	Dark matter particles inside a dark star could be annihilating into Standard Model particles, with a cross section $\langle\sigma v\rangle_{\rm ann}$, if the dark matter number is broken by two units. The rate reads
	\begin{align}
		\Gamma_{\rm ann}=\frac{1}{2}\int dV \left(\frac{\rho(r)}{m}\right)^2 \langle\sigma v\rangle_{\rm ann}.
	\end{align}
	The dark matter density was calculated in Section \ref{sec:Boson_Stars} in terms of normalized quantities. It is then convenient to define a ``normalized J-factor"
	\begin{align}
		J_*\equiv\int_{0}^{\infty}{dx_{*}\, x_{*}^2\rho_{*}(x_{*})^2},
	\end{align}
	which depends in turn on the core density. For the red, blue and green points in Fig. \ref{fig:mass_vs_x}, one finds $J_*=8.3\times 10^{-4}$, $1.3\times 10^{-4}$ and $1.0\times 10^{-5}$, respectively. This factor encompasses the differences on the choice of the particular dark star solution, and given that it doesn't have a substantial variation, we normalize it to $10^{-5}$ in what follows.
	In line with this definition, we write the annihilation rate as
	
	\begin{equation}
		\Gamma_{\rm ann}=(1.04\times10^{36}\ \mathrm{s}^{-1})\lambda^{-\frac12}\left(\frac{\langle\sigma v\rangle_{\rm ann}}{10^{-59}\,\mathrm{cm}^3\mathrm{s}^{-1}}\right)\left(\frac{J_*}{10^{-5}}\right).
	\end{equation}
	The number of DM particles in the star is dictated by
	
	\begin{equation}
		\frac{dN}{dt}=-2\Gamma_{\rm ann},
	\end{equation}
	with solution $N(t)=N_0-2\Gamma_{\rm ann}t$, where $N_0=(1.31\times10^{56})\lambda^{\frac12}\left(\frac{m}{1\ \mathrm{GeV}}\right)^{-3}$ is the initial number of particles. For simplicity, we define the lifetime of the star as the time it takes for it to lose half of its initial particles. We estimate it to be
	
	\begin{equation}
		t_{\frac12,\rm ann}=\frac{1}{4}\frac{N_0}{\Gamma_{\rm ann}}=(3.9\times10^{17}\ \mathrm{s})\lambda\left(\frac{m}{1\ \mathrm{GeV}}\right)^{-3}\left(\frac{\langle\sigma v\rangle_{\rm ann}}{10^{-59}\,\mathrm{cm}^3\mathrm{s}^{-1}}\right)^{-1},
	\end{equation}
	which leads to the constraint $\langle\sigma v\rangle_{\rm ann}<10^{-59}\,\mathrm{cm}^3\mathrm{s}^{-1}$ in order for the star to live $\sim10^{17}\ \mathrm{s}\approx3.2\ \mathrm{Gyr}$.
	The luminosity in gamma-rays generated by annihilations in the dark star can be calculated from
	\begin{align}
		L_{\rm ann}&=2m f^\gamma_{\rm ann}\Gamma_{\rm ann}
		\simeq  L_\odot f^\gamma_{\rm ann}\left(\frac{m}{1\,{\rm GeV}}\right)\lambda^{-\frac12}\left(\frac{\langle\sigma v\rangle_{\rm ann}}{1.2\times 10^{-59}\,\mathrm{cm}^3\mathrm{s}^{-1}}\right)\left(\frac{J_*}{10^{-5}}\right),
	\end{align}
	where $f^\gamma_{\rm ann}$ is the fraction of the total energy that goes into photons. Similarly, dark matter could decay, with a rate $\Gamma_{\rm dec}$, if the dark matter number is broken by one unit.  The luminosity generated in the decay reads
	\begin{align}
		L_{\rm decay}= M_{\rm DS}  f^\gamma_{\rm decay} \Gamma_{\rm dec}=(55L_\odot)f^\gamma_{\rm decay}\left(\frac{\Gamma_{\rm dec}}{10^{18}\ \mathrm{s}^{-1}}\right)^{-1}\lambda^{\frac12}\left(\frac{m}{1\ \mathrm{GeV}}\right)^{-2},
	\end{align}
	where $f_{\rm decay}$ is the fraction of decay energy that goes into photons.  Let us note that in both cases, the dark star does not lose a substantial amount of mass through annihilations or decays in cosmological timescales, so that the solution to the Einstein-Klein-Gordon equation constructed in Section \ref{sec:Boson_Stars}, which assumed a static configuration, holds. 
	
	As indicated in Section \ref{sec:Signals}, if the luminosity of the dark star is of the order of the solar luminosity, one expects the observation of the order of a million point sources in the $\gamma$-ray sky. Conversely, the non-observation of such large number of unidentified point sources translates into very stringent constraints on the annihilation cross-section. For $m\sim 1$ GeV and $\lambda\sim 1$, one obtains
	\begin{align}\label{eq:sigma_constraint}
		\langle\sigma v\rangle_{\rm ann}\lesssim 10^{-59}\,\mathrm{cm}^3\mathrm{s}^{-1}\simeq 8\times 10^{-43}\,{\rm GeV}^{-2}\simeq \frac{1.2\times 10^{-4}}{M_{\rm Pl}^2}.
	\end{align}
	It is intriguing that Planck scale suppressed annihilations could lead to observable signals. The constraint provided by Eq. (\ref{eq:sigma_constraint}) is conveniently set by the requirement that the dark star survives enough time to be present today. Even at this scale, we would expect to observe a million such objects in our Galaxy. This leads to the favorable result that, for even stronger constraints on the cross section, the possibility for observation of a dark star remains strong, as even a cross section four orders of magnitude smaller would still lead to the observation of at least one dark star, as mentioned in Section \ref{sec:Signals}.
	
	For dark matter decay, observations of the diffuse gamma-ray background lead to the bound $\Gamma^{-1}_{\rm dec}\lesssim 10^{28}\,{\rm s}$ for masses $m\sim 1\,{\rm MeV}- 10\,{\rm GeV}$. Taking for concreteness $f_{\rm decay}\sim 1/3$, the upper limit on the inverse rate translates into the upper limit on the luminosity of
	\begin{align}
		L_{\rm decay}\lesssim 10^{-9} L_\odot \lambda^{1/2}\left(\frac{m}{1\ \mathrm{GeV}}\right)^{-2}.
	\end{align}
	Using the estimate in Section \ref{sec:Signals} that for ${\cal F}_{\rm DS}=10^{-2}$ one may expect one point source in the sky if $L=10^{-4}L_\odot$. Therefore one concludes, that for $\lambda\sim1$ a dark star decaying into gamma-rays would be within experimental reach if $m\lesssim10\ \text{MeV}$.

	\section{Conclusions}
	\label{sec:conclusions}
	Dark matter with sufficiently strong self-interactions and an effective mechanism of energy evacuation can form its own compact ``dark stars". Although one would  naively expect that such stars could only be identified either by gravitational lensing observations or perhaps by detection of gravitational waves from merger events that cannot be attributed neither to black holes nor to neutron stars, surprisingly enough capture of interstellar protons and electrons by such dark stars could lead to photon emission outbursts with luminosities depending on the parameters up to several orders of magnitude larger than the solar one. Under certain conditions the protons and electrons once captured, settle at the core of the star within a thermal radius where they can emit gamma rays via bremsstrahlung interactions. Eventually such interactions dominate the cooling rate of the star. As the star cools down, the thermal radius shrinks leading to a culminating emissivity that can cause an outburst which gives a clear distinctive and characteristic signal that could easily be detected in sky, thus providing a smoking gun identification signature for the existence of such ``dark stars" in the Universe.
	
	\acknowledgments
	
	This work was supported by the Collaborative Research Center SFB1258 and by the Deutsche Forschungsgemeinschaft (DFG, German Research Foundation) under Germany's Excellence
	Strategy - EXC-2094 - 390783311.

	\bibliographystyle{JHEP} % We choose the "plain" reference style
	\bibliography{References} % Entries are in the References.bib file
	
\end{document}